\documentclass[%
 reprint,
 amsmath,amssymb,
 aps,
prb,
]{revtex4-2}

\usepackage{graphicx}
\usepackage{dcolumn}% Align table columns on decimal point
\usepackage{bm}% bold math
\begin{document}

%\preprint{APS/123-QED}

\title{Unsupervised learning-based structural analysis: Search for a characteristic low-dimensional space by local structures in atomistic simulations}% Force line breaks with \\

\author{Ryo Tamura}
\email{tamura.ryo@nims.go.jp}
\affiliation{International Center for Materials Nanoarchitectonics, National Institute for Materials Science, Tsukuba 305-0044, Japan}
\affiliation{Research and Services Division of Materials Data and Integrated System, National Institute for Materials Science, Tsukuba 305-0047, Japan}
\affiliation{Graduate School of Frontier Sciences, The University of Tokyo, Chiba 277-8568, Japan}
 
\author{Momo Matsuda}
\affiliation{Department of Computer Science, University of Tsukuba, Tsukuba 305-8573, Japan}

\author{Jianbo Lin}
\affiliation{International Center for Materials Nanoarchitectonics, National Institute for Materials Science, Tsukuba 305-0044, Japan}
\affiliation{Center for Artificial Intelligence, University of Tsukuba, Tsukuba 305-8573, Japan}

\author{Yasunori Futamura}
\affiliation{Department of Computer Science, University of Tsukuba, Tsukuba 305-8573, Japan}
\affiliation{Center for Artificial Intelligence, University of Tsukuba, Tsukuba 305-8573, Japan}
\affiliation{Master's/Doctoral Program in Life Science Innovation, University of Tsukuba, Tsukuba 305-8577, Japan}

\author{Tetsuya Sakurai}
\email{sakurai@cs.tsukuba.ac.jp}
\affiliation{Department of Computer Science, University of Tsukuba, Tsukuba 305-8573, Japan}
\affiliation{Center for Artificial Intelligence, University of Tsukuba, Tsukuba 305-8573, Japan}
\affiliation{Master's/Doctoral Program in Life Science Innovation, University of Tsukuba, Tsukuba 305-8577, Japan}

\author{Tsuyoshi Miyazaki}
\email{miyazaki.tsuyoshi@nims.go.jp}
\affiliation{International Center for Materials Nanoarchitectonics, National Institute for Materials Science, Tsukuba 305-0044, Japan}
\affiliation{Master's/Doctoral Program in Life Science Innovation, University of Tsukuba, Tsukuba 305-8577, Japan}

\date{\today}% It is always \today, today,

\begin{abstract}
Owing to the advances in computational techniques and the increase in computational power, 
atomistic simulations of materials can simulate large systems with higher accuracy. 
Complex phenomena can be observed in such state-of-the-art atomistic simulations. However, it has become increasingly difficult to understand what is actually happening and mechanisms, for example, in molecular dynamics (MD) simulations. 
We propose an unsupervised machine learning method to analyze the local structure around a target atom. 
The proposed method, which uses the two-step locality preserving projections (TS-LPP), can find a low-dimensional space wherein the distributions of datapoints for each atom or groups of atoms can be properly captured.
We demonstrate that the method is effective for analyzing the MD simulations of crystalline, liquid, and amorphous states and the melt-quench process from the perspective of local structures.
The proposed method is demonstrated on a silicon single-component system, a silicon-germanium binary system, and a copper single-component system.
\end{abstract}

\maketitle

%\tableofcontents

%%%%%%%%%%%%%%%%%%%%%%%
%%%%%%%%%%%%%%%%%%%%%%%
\section{\label{sec:intro}Introduction}
%%%%%%%%%%%%%%%%%%%%%%%

The recent progress in atomistic simulations is remarkable. 
This approach can be implemented to simulate large systems with higher accuracy with the help of advanced computational techniques and improved computational power. 
Compared with the classical molecular dynamics (MD) simulation, which uses empirical atomic force fields, the MD simulation based on the density functional theory (DFT), which is called first-principles MD (FPMD), is reliable even for new materials or unknown phases considering the lack of experimental information~\cite{car_unified_1985,car_structural_1988,blochl_adiabaticity_1992,pasquarello_interface_1998,oganov_elastic_2001,hirata_direct_2011,sang_situ_2018,reocreux_reactivity_2019,shi_aqueous_2019}. 
With the progress in large-scale DFT calculation techniques~\cite{bowler_on_2012}, FPMD simulations for larger system sizes can be realized. 
Machine-learning (ML) techniques can construct atomic force fields, which can reproduce the DFT results quite accurately~\cite{behler_generalized_2007,behler_atom-centered_2011,pham_novel_2016,bartok_machine_2017,suzuki_machine_2017,chen_accurate_2017,kobayashi_neural_2017,tamura_machine_2019,li_dependence_2019}. 
With such ML force fields, the computation time is greatly reduced, 
and long-term MD simulations of large systems are possible~\cite{li_molecular_2015,deringer_realistic_2018}.

As recent FPMD simulations can be utilized for explaining various phenomena and complex structural changes, it has become increasingly difficult to understand what is actually happening during such simulations. 
In some cases, a local or global unknown phase or structure is encountered. 
Even well-known materials may have a hidden structural order in the middle range~\cite{stillinger_hidden_1982,lee_hidden_2013,tong_revealing_2018}. 
To analyze the structural properties of materials, radial distribution functions (RDFs) or pair correlation functions have been traditionally used~\cite{mcgreevy_reverse_1988,li_radial_1990,lacevic_growing_2002}. 
Recently, analysis based on the persistent homology diagram has been employed~\cite{hiraoka_hierarchical_2016,saadatfar_pore_2017,onodera_understanding_2019}. 
However, these methods require statistics over many atoms or from a long-time profile. 
For FPMD simulations with a large system size, the system may not be in a single phase or can be irregular. 
Thus, it is desirable to develop an analysis method with small statistics based on the local structure of the atoms.

To solve this problem, some methods for local structure detection have been proposed~\cite{lechner_accurate_2008,geiger_neural_2013,piaggi_entropy_2017,caro_reactivity_2018,boattini_neural-network-based_2018,boattini_unsupervised_2019,boattini_autonomously_2020}. 
A common feature is that a local structure around each atom in a snapshot is expressed by a relatively high-dimensional descriptor (high-dimensional vector).
The bond order parameter, smooth overlap of atomic positions (SOAP), and  atom-centered symmetry functions (ACSF) are typically used as descriptors. 
The datasets, where the descriptors for many atoms are collected, are analyzed using supervised or unsupervised ML methods, such as neural networks to detect the differences in the local structures for each atom.
These methods have been used to detect differences in local structures of crystal~\cite{lechner_accurate_2008}, amorphous~\cite{caro_reactivity_2018}, and supercooled liquid compounds~\cite{boattini_autonomously_2020}.

In this study, we propose an unsupervised learning-based method to analyze local structures with the same motivation as that for previous studies.
Specifically, the method aims to find a low-dimensional space, 
where the distributions of datapoints for each atom or groups of atoms can be accurately captured.
The major difference between the proposed method and previous methods for local structure detection is that we use locality preserving projections (LPP) to perform dimensionality reduction. 
This allows us to obtain a linear transformation from a high-dimensional descriptor to a low-dimensional space, similar to that achieved through principal component analysis (PCA).
The advantage to use linear transformation is that the exact linear transformation of new data points (out-of-samples) can be achieved.
In addition, by a linear transformation,
data mapping to low dimensional space, 
which reflects the original data structure in the high dimensional space,
can be realized.
Note that a non-linear transformation may significantly change the original data structure in the reduced low-dimensional space.

\begin{figure}[t]
\centering
\includegraphics[scale=0.75]{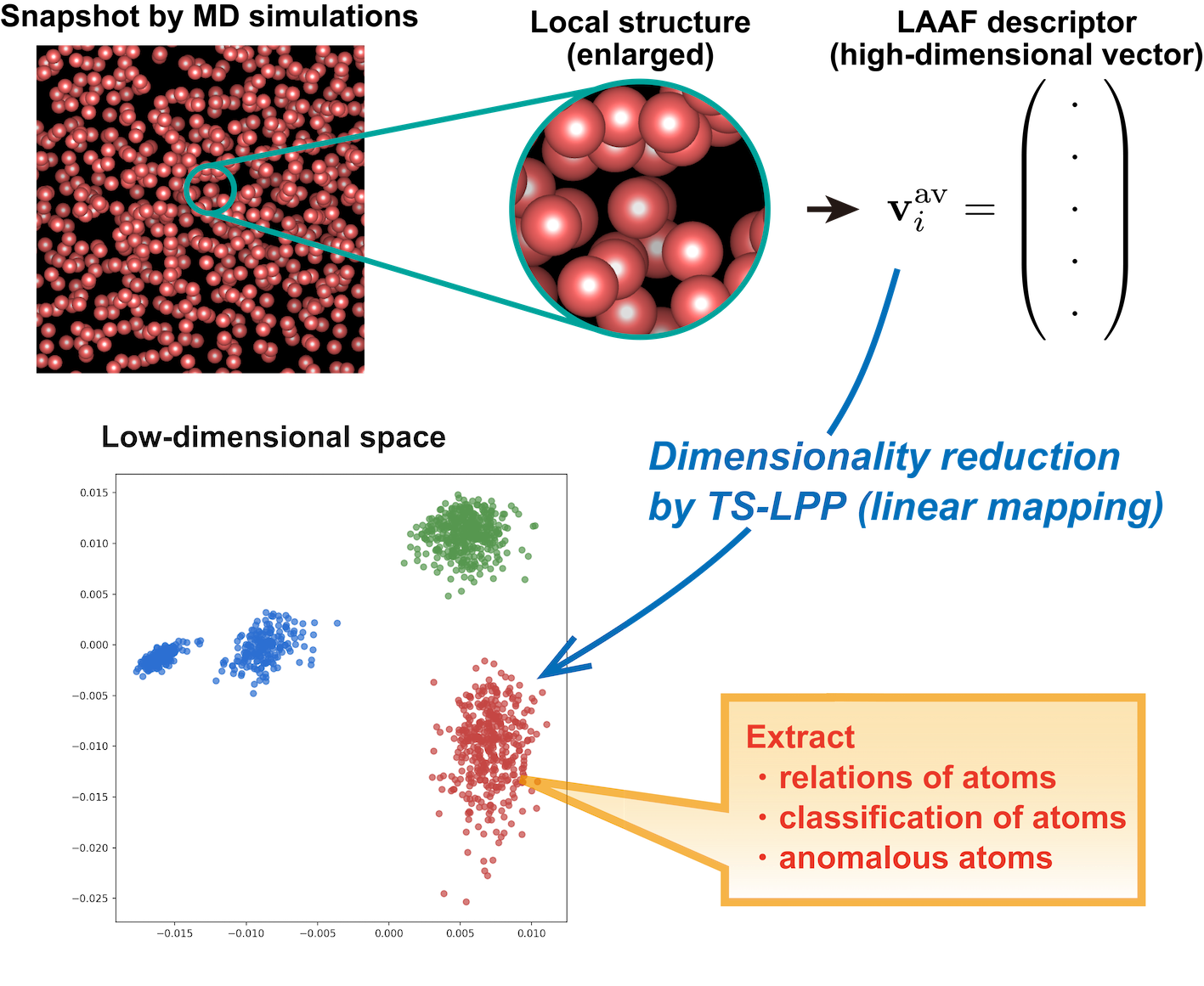}
 \caption{
Schematic of unsupervised learning based structural analysis.
}
\label{flow}
\end{figure}

The proposed method has three steps, as shown in Fig.~\ref{flow} (detailed in Sec.~\ref{sec:methods}). 
Firstly, a high-dimensional descriptor, which is called as locally averaged atomic fingerprints(LAAF) descriptor, is introduced to represent the local structure.
The LAAF descriptor shows a two-body correlation for the target atom and its neighboring atoms. 
Here, the LAAF descriptor using two-body correlation is the simplest descriptor belonging to the ACSF. 
Although many better descriptors have been developed, 
we adopt the simplest descriptor to verify the usefulness of our dimensionality reduction method.
Secondly, a dimensionality reduction scheme is employed. 
As a dimensionality reduction method, the two-step locality preserving projections (TS-LPP) method, in which the calculations by conventional LPP~\cite{he_locality_2004} are repeated two times, is proposed. 
It is important that the dimensionality reduction is unsupervised because it should detect the atoms in a new or unknown phase. 
Moreover, dimensionality reduction will ensure that a more reliable measure is introduced to evaluate the similarity between the two data. 
In the low-dimensional space, the classification of new data points (i.e., out-of-sample points) is expected to be more robust.
Thirdly, in the low-dimensional space,
we extract relations of each atom in terms of the local structure,
perform the classification of new data points,
and detect singular atoms which characterize the structure and drive phase transitions.

To validate the proposed unsupervised learning-based structural analysis method, we introduce the results of some examples.
First, we perform unsupervised clustering of the local structure data sampled from snapshots of silicon in the crystalline, liquid, and amorphous phases in Sec.~\ref{sec:Si_single}. 
To realize the unsupervised method, the information of the detected phases during dimensionality reduction is not used. 
The widely used PCA method cannot find a low-dimensional space where the three phases are clearly divided using their local structures. 
In contrast, in a low dimensional space obtained by TS-LPP method, three well-separated groups, each of which  corresponds to its correct phase, are obtained. 
Furthermore, we discuss why TS-LPP works well in comparison to PCA and LPP.
Next, we focus on the structural analysis of the melt-quench process from liquid to amorphous formation in the silicon system in Sec.~\ref{sec:melt-quench}.
Since linear mapping function for datapoints is available with the TS-LPP method,
we can analyze new datapoints in the low-dimensional space.
Thus, each atom in the snapshots from the melt-quench process is plotted and analyzed in a low-dimensional space.
We address the behavior of the atoms in the process and the amorphous structures depending on the melt-quench processes.
In addition, the silicon-germanium binary system and the copper system are targeted for the proposed unsupervised learning-based structural analysis in Secs.~\ref{sec:SiGe_binary} and \ref{sec:Cu_single}.
Discussion and Summary are given in Sec. IV.

%%%%%%%%%%%%%%%%%%%%%%%
%%%%%%%%%%%%%%%%%%%%%%%
\section{\label{sec:methods}Methods}
%%%%%%%%%%%%%%%%%%%%%%%

In this section, we explain an unsupervised learning-based method to analyze the local structure.
The proposed structural analysis method uses the LAAF descriptor to express the local structure of each atom as high-dimensional vectors and the TS-LPP as the dimensionality reduction method.
In addition,
the setups of the FPMD and classical MD methods to generate snapshots of MD simulations are introduced.

%%%%%%%%%
\subsection{LAAF descriptor}

To express the local structure of the target atom using a high-dimensional vector, we use the atomic fingerprint proposed by Botu and Ramprasad~\cite{botu_adaptive_2015}. 
For single-component systems, the atomic fingerprint vector of the $i$th atom is written as
\begin{eqnarray}
V_i (\eta_m; R_{\rm c}) &=& \sum_{j \neq i} \exp \left[ - (r_{ij}/\eta_m)^2 \right] f(r_{ij}; R_{\rm c}), \\
f(r_{ij}; R_{\rm c}) &=& 
\begin{cases} 
0.5 \left[ \cos (\pi r_{ij}/R_{\rm c})^2 \right] + 1 & (r_{ij} \le R_{\rm c}) \\
0 & (r_{ij} > R_{\rm c})
\end{cases},
\end{eqnarray}
where $f(r_{ij};R_{\rm c})$ is the cutoff function with cutoff $R_{\rm c}$, which is related to the locality of the structural order in the two-body distribution function. 
Furthermore, $r_{ij}=|\mathbf{r}_{i} - \mathbf{r}_{j}|$ is the distance between the position $\mathbf{r}_{i}$ of the $i$th atom and the position $\mathbf{r}_{j}$ of the $j$th neighbor atom, and $\eta_{m}  (m=1,... ,M)$ is the decay rate with distance. 
Thus, an $M$-dimensional atomic fingerprint vector for $i$th atom is obtained as
\begin{eqnarray}
\mathbf{V}_i (R_{\rm c}) = \left( V_i (\eta_1; R_{\rm c}), V_i (\eta_2; R_{\rm c}), ..., V_i (\eta_M; R_{\rm c}) \right).
\end{eqnarray}

The LAAF descriptor with cutoff $R_{\rm a}$ from $i$th atom corresponds to the locality with respect to the statistics (similarity) around the target atom and is calculated by
\begin{eqnarray}
\mathbf{V}_i^{\rm av} (R_{\rm c}, R_{\rm a}) = \frac{1}{N_{\in R_{\rm a}}} \sum_{j \in R_{\rm a}} \mathbf{V}_j (R_{\rm c}),
\end{eqnarray}
where $N_{\in R_{\rm a}}$ is the number of atoms within the average radius $R_{\rm a}$ from the $i$th atom, and the sum is computed for the $j$th atom within $R_{\rm a}$ from the $i$th atom. 
If $R_{\rm a}$ and $R_{\rm c}$ are sufficiently small, the LAAF descriptor expresses the local structure around the $i$th atom. 
As $R_{\rm a}$ and $R_{\rm c}$ increase, locality decreases. 
In this study, we set $M=100$, and $\eta_m$ is defined by a logarithmic grid up to $R_{\rm c}$.

In addition, the LAAF descriptor for binary systems can be defined in a straightforward manner.
We consider a system consisting of two elements, $a$ and $b$.
The atomic fingerprint vectors of the $i$th atom of element $a$ are defined as
\begin{eqnarray}
V^a_{i \in a} (\eta_m; R_{\rm c}) &=& \sum_{j \in a \text{.and.} j \neq i} \exp \left[ - (r_{ij}/\eta_m)^2 \right] f(r_{ij}; R_{\rm c}), \notag \\ \\
V^b_{i \in a} (\eta_m; R_{\rm c}) &=& \sum_{j \in b} \exp \left[ - (r_{ij}/\eta_m)^2 \right] f(r_{ij}; R_{\rm c}),
\end{eqnarray}
where the former and latter equations capture the structure of elements $a$ and $b$ around the $i$th atom, respectively.
When we use $\eta_{m}  (m=1,... ,M)$,
a $2M$-dimensional atomic fingerprint vector for the $i$th $a$ element atom is obtained as
\begin{eqnarray}
\mathbf{V}_{i \in a} (R_{\rm c}) &=& ( V^a_{i \in a} (\eta_1; R_{\rm c}), V^a_{i \in a} (\eta_2; R_{\rm c}), ..., V^a_{i \in a} (\eta_M; R_{\rm c}), \notag \\
&&V^b_{i \in a} (\eta_1; R_{\rm c}), V^b_{i \in a} (\eta_2; R_{\rm c}), ..., V^b_{i \in a} (\eta_M; R_{\rm c})). \notag \\
\end{eqnarray}
Using the cutoff $R_{\rm a}$, the LAAF descriptor for the $a$-element atom can be calculated as
\begin{eqnarray}
\mathbf{V}_{i \in a}^{\rm av} (R_{\rm c}, R_{\rm a}) = \frac{1}{N_{a, \in R_{\rm a}}} \sum_{j \in R_{\rm a}} \mathbf{V}_{j \in a} (R_{\rm c}),
\end{eqnarray}
where $N_{a, \in R_{\rm a}}$ is the number of $a$-element atoms within the average radius $R_{\rm a}$ from the $i$th atom 
and the sum is computed for the $j$th $a$-element atom within $R_{\rm a}$ from the $i$th atom. 
To calculate the LAAF descriptor for the $b$-element atom,
$a$ is swapped with $b$ in the equations.

%%%%%%%%%
\subsection{TS-LPP method}

\begin{figure*}[t]
\centering
\includegraphics[width=1.\textwidth]{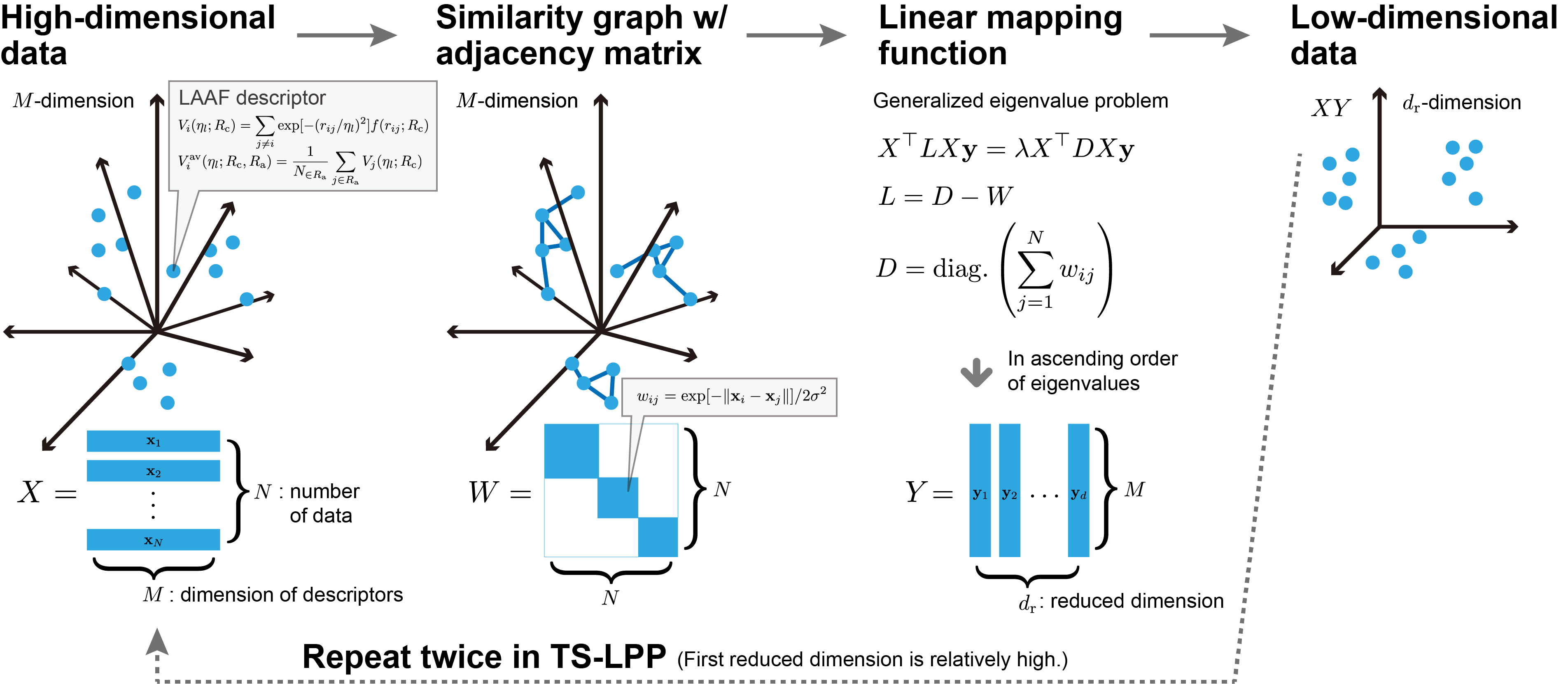}
 \caption{
Procedure of the two-step locality preserving projection (TS-LPP) method.
The input data matrix $X$ is constructed by the $M$-dimensional feature vector $\mathbf{x}_i$ defined by the LAAF descriptor sampled from molecular dynamics simulations. 
Through a similarity graph with an adjacency matrix $W$, similar data are embedded to the closer positions in low-dimensional space, 
and a linear mapping matrix $Y$ is obtained by solving the generalized eigenvalue problem. 
In TS-LPP, first, the embedded data in the middle-dimensional space, which is larger than the target dimension, is calculated. 
Subsequently, dimension reduction to the final target dimension is performed again. 
In the low-dimensional space, we extract the relations of each atom in terms of the local structure using the TS-LPP method.
}
\label{LPP_fig}
\end{figure*}

We propose the TS-LPP method as a dimensionality reduction method and an automated determination technique for the hyperparameters. 
The flow of this algorithm is shown in Fig.~\ref{LPP_fig}.
Consider the case in which the dimension of the LAAF descriptors is $M$ and the dimension of the target low-dimensional space is $d_{\rm r}$. 
In TS-LPP, the initial dimension reduction from $M$-dimensional space to $d_{\rm m}$-dimensional space ($d_{\rm r} < d_{\rm m} < M$) is performed by conventional LPP~\cite{he_locality_2004}. 

The LPP method is as follows (see Fig.~\ref{LPP_fig}). 
The input data matrix $X$ is composed of $M$-dimensional feature vectors $\{\mathbf{x}_i \}_{i=1,... ,N}$, where $N$ is the number of data points. 
In this study, $\mathbf{x}_i$ is the $i$th LAAF descriptor, which is standardized by the z-score. 
The method begins with a calculation of the weighted adjacency matrix $W$ where each component is defined as follows:
\begin{eqnarray}
W_{ij} = \begin{cases}
\exp \left[ - \| \mathbf{x}_i - \mathbf{x}_j \| \right] / 2 \sigma^2 & (i \neq j) \\
0 & (i=j)
\end{cases},
\end{eqnarray}
where $\sigma$ is a hyperparameter, and this adjacency matrix is the $N \times N$ matrix. 
To create the $k_{\rm nn}$-nearest neighbor graph (similarity graph), for each column in $W$, 
the off-diagonal elements, except those related to the $k_{\rm nn}$-nearest neighbor data points, are forced to 0. 
Here, we fix $k_{\rm nn}=7$. 
Next, to represent the adjacency matrix as a symmetric matrix, the following operation is performed: $W_{ij} = \max \left( W_{ij}, W_{ji} \right)$~\cite{Luxburg-2007}.
Furthermore, we introduce the degree matrix $D$, 
which is the $N \times N$ diagonal matrix given as
\begin{eqnarray}
D = \begin{pmatrix} d_1 & \cdots & 0 \\ \vdots & \ddots & \vdots \\ 0 & \cdots & d_N \end{pmatrix},
\end{eqnarray}
where each diagonal component is defined as
\begin{eqnarray}
d_i = \sum_{j=1}^N W_{ij}.
\end{eqnarray}
Using $W$ and $D$, the graph Laplacian is defined as
\begin{eqnarray}
L = D - W.
\end{eqnarray}
In LPP, we solve the following generalized eigenvalue problem
\begin{eqnarray}
X^\top L X \mathbf{y} = \lambda X^\top D X \mathbf{y}. \label{eq:LPP}
\end{eqnarray}
Here, $\{\lambda_i, \mathbf{y}_i \}$ is the $i$th eigenvalue and eigenvector of this problem, which are arranged in ascending order.

Let $d_{\rm r}$ denote the target dimension for dimension reduction. 
Then, the mapping matrix by LPP from $M$-dimensional space to $d_{\rm r}$-dimensional space is given by
\begin{eqnarray}
Y= \left( \mathbf{y}_1, \mathbf{y}_2,..., \mathbf{y}_{d_{\rm r}} \right).
\end{eqnarray}
Using this mapping matrix, the feature vector $\mathbf{x}$ is mapped to the low-dimensional vector $\mathbf{x}'$ with $d_{\rm r}$ dimensions as
\begin{eqnarray}
\mathbf{x}' = \mathbf{x} Y.
\end{eqnarray}
In the numerical calculation for obtaining $Y$,
the low-rank approximation is used to improve the numerical stability,
which is explained in Appendix A.

In TS-LPP, we conduct LPP from $d_{\rm m}$-dimensional space to $d_{\rm r}$-dimensional space. 
Thus, $d_{\rm m}$ is a hyperparameter in this method. 
Furthermore, to perform LPP, another hyperparameter $\sigma$ exists to prepare the weighted adjacency matrix. 
To determine the appropriate hyperparameters $d_{\rm m}$ and $\sigma$, the Calinski-Harabasz score (Pseudo-F)~\cite{calinski_dendrite_1974} is used. 
This score is defined as the ratio between the intra-cluster and inter-cluster dispersions. 
This score is large when the distances between clusters are large, and each cluster is dense. 
A grid search for hyperparameters $d_{\rm m}$ and $\sigma$ is performed to maximize the Calinski-Harabasz score, using scikit-learn~\cite{noauthor_httpsscikit-learnorgstablemodulesgeneratedsklearnmetricscalinski_harabasz_scorehtml_2020}. 
To evaluate the score, the labels obtained by the k-means method~\cite{noauthor_httpsscikit-learnorgstablemodulesgeneratedsklearnclusterkmeanshtml_2020} with a fixed cluster number are used. 
Thus,
it is necessary to analyze the cluster number dependency of the score.
However, as shown in the results, 
this dependency is not critical, and the cluster number is not an important factor. 
Therefore, we try several values of cluster numbers and adopt the cluster number with which the distributions of datapoints for each atom or groups of atoms can be properly captured in the low-dimensional space.
The scheme mentioned above is unsupervised because information about the correct state of the training data is not used. 
Note that the common value of $\sigma$ is used in the first and second LPP calculations.

%%%%%%%%%
\subsection{Characteristic of TS-LPP method} \label{sec:synthetic_TS-LPP}

\begin{figure*}[t]
\centering
\includegraphics[scale=0.55]{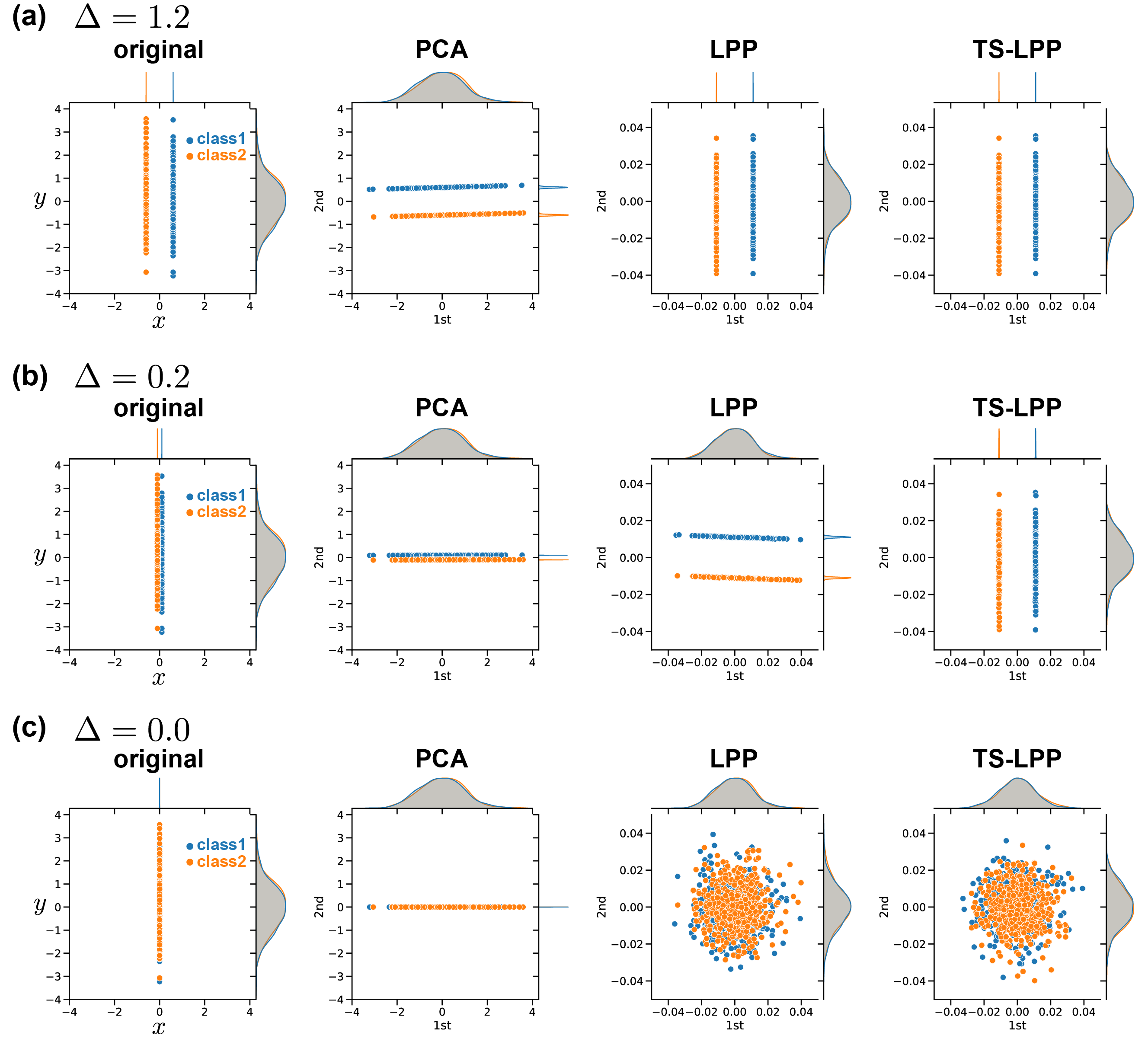}
\caption{
Two dimensional synthetic data where the two classes are widely distributed along the $y$-axis and the distance between the classes in the $x$-axis is gradually decreased as (a) 1.2, (b) 0.2, and (c) 0.0.
The original data, PCA, LPP, and TS-LPP results are shown.
}
\label{2D_discuss_artificial}
\end{figure*}

To demonstrate the advantage of TS-LPP method, 
we introduce an example when a 2D synthetic dataset is widely distributed along the $y$-axis and the distance $\Delta$ between the two classes in the $x$-axis is gradually decreased.
The prepared synthetic dataset is shown in Fig.~\ref{2D_discuss_artificial}.
As we discuss later, similar data structure is observed in the distribution of the data generated by the MD simulations in this work.
This 2D synthetic dataset is randomly generated from the normal distribution with mean of 0 and standard deviation of $10^{-6}$ for $x$-axis and 0.98 for $y$-axis, respectively.
Each class has 500 datapoints and the distance between these classes along $x$-axis is changed as $\Delta = 1.2$, 0.2, and 0.0.
In this demonstration, hyperparameter $\sigma$ is fixed as 5.
In Fig.~\ref{2D_discuss_artificial} (a), we show scatter plots in the case of $\Delta = 1.2$.
Using dimensionality reduction methods, we intend to obtain 1D space that allow one to (linearly) separate class 1 (orange) and 2 (blue).
In the panels representing PCA, LPP, and TSLPP results, we show not only the first principal axis (shown as $x$-axis) derived by a dimensionality reduction but also the second axis ($y$-axis) in order to see the behaviors of the linear maps.
As shown in the left most panel (Original) of the figure, class 1 and 2 can be separated on $x$-axis.
The most widespread $y$-axis cannot be used to distinguish class 1 and 2.
This data distribution makes the first axis of PCA which fails to separate class 1 and 2.
In contrast, both LPP and TS-LPP are successful to determine the first axis that can separate the classes.

Furthermore, the results of LPP and TS-LPP are different when the distance between two classes is smaller ($\Delta = 0.2$) as shown in Fig.~\ref{2D_discuss_artificial} (b).
While the first axis of LPP cannot separate the classes as with PCA, the distance between the classes on the second axis is larger.
TS-LPP applies LPP again to the 2D space obtained by LPP.
The resulting first axis successfully separates the classes.
Note that even if LPP is performed after PCA, the classes are not separated in the principal axis.
We also show the case of $\Delta = 0$ in Fig.~\ref{2D_discuss_artificial} (c).
In this case, the distance between class 1 and 2 along the $x$-axis is zero,
and all methods unsurprisingly fail to separate the classes.
For PCA, the most widespread $y$-axis is chosen as the first axis,
which is also reasonable. 

This validation on synthetic data shows that TS-LPP performs well when the intra-class variation is large and the inter-class distance is fairly small.
In addition, if the distance between the classes is large enough, 
the data can be separated even with a single LPP. 
These results can be explained by the locality preserving nature of LPP. 
In LPP, close data in the original space are mapped to close positions in the low-dimensional space. 
In other words, if the distance between two classes is too small, 
the data will be mapped to close positions in the first axis, 
and the two datasets will be mixed in the low-dimensional space. 
On the other hand, 
even if the distance between the two classes is not large in the original space,
this distance is extended in the second axis by LPP.
This situation is not observed by PCA, and thus this is an advantage to use LPP.
As will be demonstrated in Sec.~\ref{sec:results},
this aspect of LPP or TS-LPP is important to find out the subtle differences between the similar data of local structures in different phases.

%%%%%%%%%
\subsection{MD simulations}

To generate MD simulation results for silicon single-component system and silicon-germanium binary system, we perform FPMD simulations using a linear-scaling method with the CONQUEST code~\cite{noauthor_http://www.order-n.org/_2018,Nakata-2020}. 
The accuracy of this calculation method is explained in Refs. ~\cite{bowler_recent_2002,miyazaki_atomic_2004,bowler_recent_2006}.
The local density approximation with the standard Ceperley--Alder exchange-correlation functional is used. 
Using the Siesta code~\cite{soler_siesta_2002}, Troullier--Martins-type norm-conserving pseudopotentials and the pseudo-atomic orbital basis sets are generated. 
We employ a minimal basis set with a cutoff energy for the charge density grid of 80 Hartree. 
The density matrix minimization (DMM) method realizes linear-scaling FPMD simulations~\cite{arita_stable_2014}. 
The cutoff range of the auxiliary density matrix in the DMM method is 16.0 Bohr. 
Using the Nose--Hoover chain thermostats, canonical ensemble (NVT) simulations~\cite{hirakawa_canonical-ensemble_2017} are carried out.

For the copper single-component system,
classical MD simulations with an effective medium theory (EMT) potential are performed using the atomic simulation environment (ASE) package~\cite{larsen_atomic_2017}.
We prepare initial structures for crystalline or liquid states with the Maxwell-Boltzmann distribution of the velocity at the targeted temperatures. 
NVT-MD simulations are conducted using the time step of 1 fs and Langevin thermostat with a friction value of 0.01 atomic unit.

%%%%%%%%%%%%%%%%%%%%%%%
%%%%%%%%%%%%%%%%%%%%%%%
\section{\label{sec:results}Results}
%%%%%%%%%%%%%%%%%%%%%%%

To show a potential of our method,
the results of the proposed unsupervised structural analysis are presented in this section.
The target systems are the silicon single-component system, silicon and germanium binary system prepared by FPMD simulations, and the copper single-component system prepared by classical MD simulations.

%%%%%%%%%
\subsection{\label{sec:Si_single} Structural analysis of crystalline, liquid, and amorphous states in Si system}

To verify the efficiency of the proposed unsupervised structural analysis method,
we work on a silicon system in the crystalline, liquid, and amorphous phases as the simplest case.
We show that TS-LPP can find the low-dimensional space where the three states are completely divided even if $R_{\rm a}$ and $R_{\rm c}$ values are small.

\begin{figure}[t]
\centering
\includegraphics[scale=0.22]{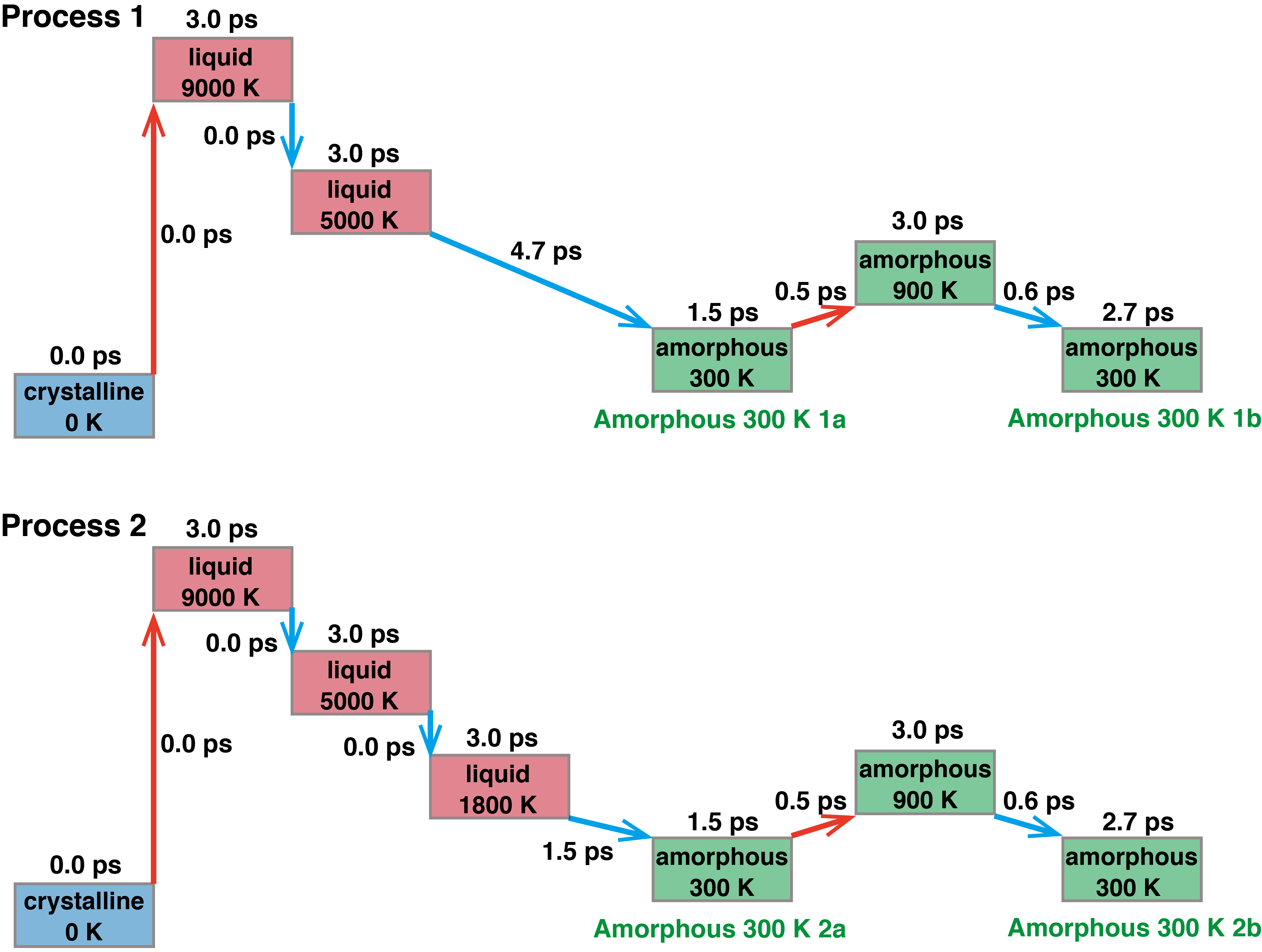}
 \caption{
Two melt-quench processes to prepare the amorphous states.
}
\label{process_amo}
\end{figure}

The various structures in the three states are prepared by the FPMD simulations for a system containing 1000 silicon atoms at constant temperature and constant volume. 
FPMD simulations are performed for the crystalline states at 300 K and 1200 K; liquid states at 3000 K, 5000 K, and 9000 K; and the four amorphous states labeled as 1a, 1b, 2a, and 2b at 300 K. 
The amorphous states are prepared using two different melt-quench processes, as shown in Fig.~\ref{process_amo}.
The total time of the FPMD simulations is 19 ps in both cases. 
The RDFs for these states are shown in Fig.~S1,
and these results are consistent with those previously reported~\cite{ishimaru_atomistic_2001,alvarez_radial_2002,ganesh_liquid-liquid_2009,wang_nanometric_2017,bartok_machine_2018}.

%%%%
\subsubsection{Distributions of datapoints in the 2D spaces obtained by TS-LPP, LPP, and PCA}

\begin{figure*}[t]
\centering
\includegraphics[scale=0.45]{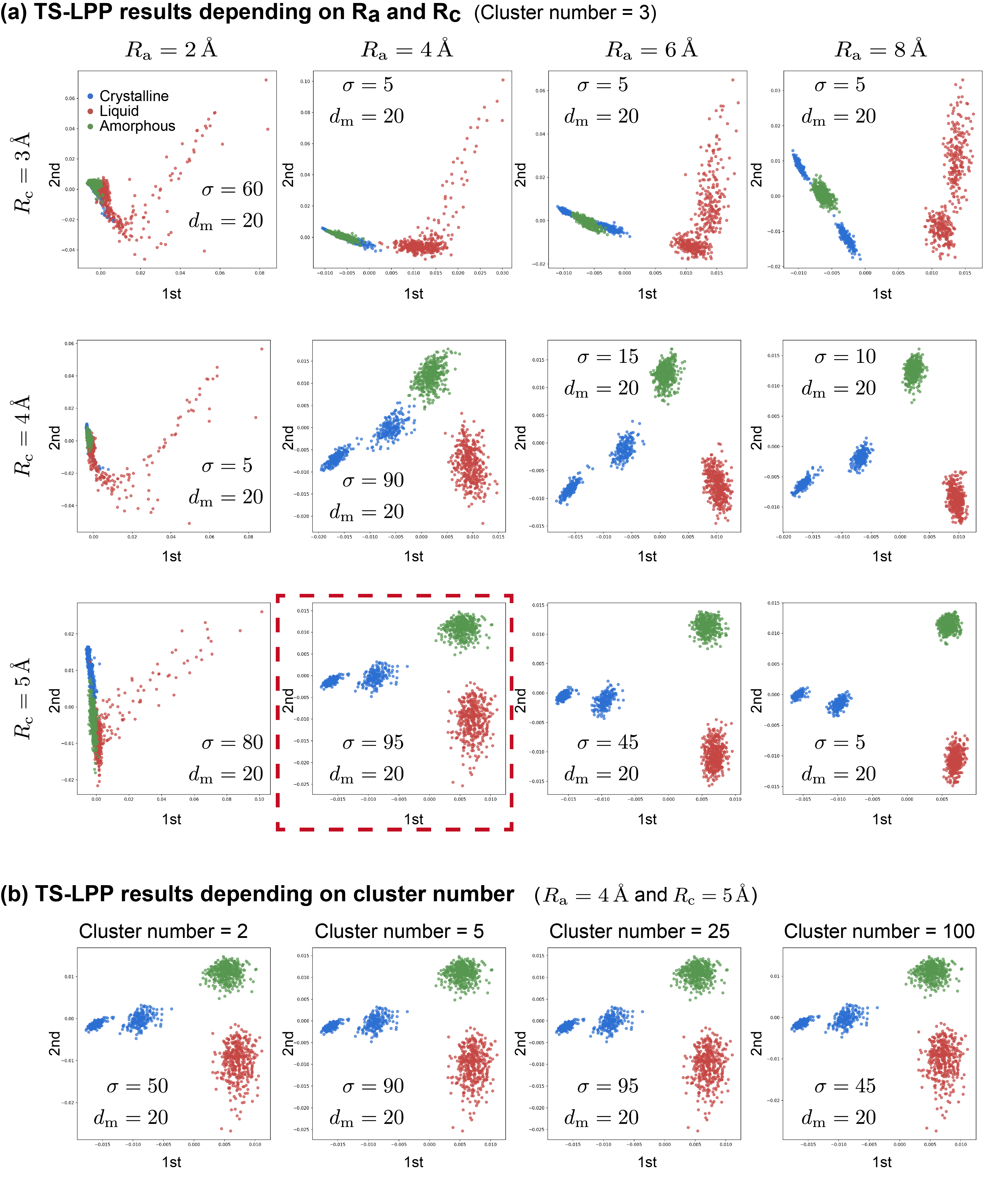}
 \caption{
(a) Distributions after dimension reduction to $d_{\rm r}=2$ by the TS-LPP method depending on $R_{\rm a}$ and $R_{\rm c}$.
Blue, red, and green indicate crystalline, liquid, and amorphous states, respectively.
The values of the hyperparameters $\sigma$ and $r_m$ are determined by the Calinski-Harabasz score when the cluster value is fixed at three.
When $R_{\rm a} = 4 {\rm \AA}$ and $R_{\rm c} = 5 {\rm \AA}$, perfect clustering is obtained, and the result is surrounded by a red dotted line.
(b) Cluster number dependence on the distributions and hyperparameters when  $R_{\rm a} = 4 {\rm \AA}$ and $R_{\rm c} = 5 {\rm \AA}$ by the TS-LPP method.
The results are almost the same with as in the three-cluster case.
}
\label{2d_three_states}
\end{figure*}

To find a low-dimensional space,
we prepare a target dataset including atoms expressed by the LAAF descriptor in the crystalline states at 300 K and 1200 K, the liquid states at 3000 K and 9000 K, and the amorphous states 1a and 1b at 300 K.
The number of datapoints in each state is 200,
and the total number of data is 1200.
Figure~\ref{2d_three_states} (a) shows the distributions in 2D space ($d_{\rm r} = 2$) obtained by the TS-LPP method depending on two parameters $R_{\rm a}$ and $R_{\rm c}$.
$R_{\rm a}$ is related to the locality of the structural order in the two-body distribution function,
and $R_{\rm c}$ corresponds to the locality with respect to the statistics (similarity) around the target atom.
Thus, local structure analysis can be performed when these parameters are sufficiently small.
Note that, to determine the appropriate hyperparameters,
the cluster number is fixed as three, 
because we assume that the data will be grouped into three clusters.
However, as will be shown later,
this value has little effects on the results.

By increasing $R_{\rm a}$ and $R_{\rm c}$,
the distributions of datapoints for the three states are distinguished.
When $R_{\rm a} = 4 {\rm \AA}$ and $R_{\rm c} = 4 {\rm \AA}$,
the three distributions are well distinguished, although a small overlap between the crystalline and amorphous states is observed.
The case of $R_{\rm a} = 4 {\rm \AA}$ and $R_{\rm c} = 5 {\rm \AA}$ is the best result when the three states are completely divided.
The typical interatomic distance between Si atoms is 2.2 ${\rm \AA}$ for the nearest and 3.8 ${\rm \AA}$ for the next-nearest neighbor atoms. 
This information is included in the LAAF descriptor when $R_{\rm c}=5 {\rm \AA}$ at least. 
This means that when the LAAF descriptor of a target atom is analyzed, if the atomic fingerprints for the nearest neighbor and next-nearest neighbor atoms are also considered, 
the phase of the target atom can be identified perfectly. 

We consider the results depending on the number of clusters for determining hyperparameters. 
Figure~\ref{2d_three_states} (b) shows the distributions in the reduced two dimensions for different cluster numbers. 
The values of the optimized hyper-parameters are also provided. 
It is encouraging that the results are almost the same. 
This suggests the present analysis method can extract a useful low dimensional space to classify the local structures without any preceding information.

\begin{figure}[t]
\centering
\includegraphics[scale=0.5]{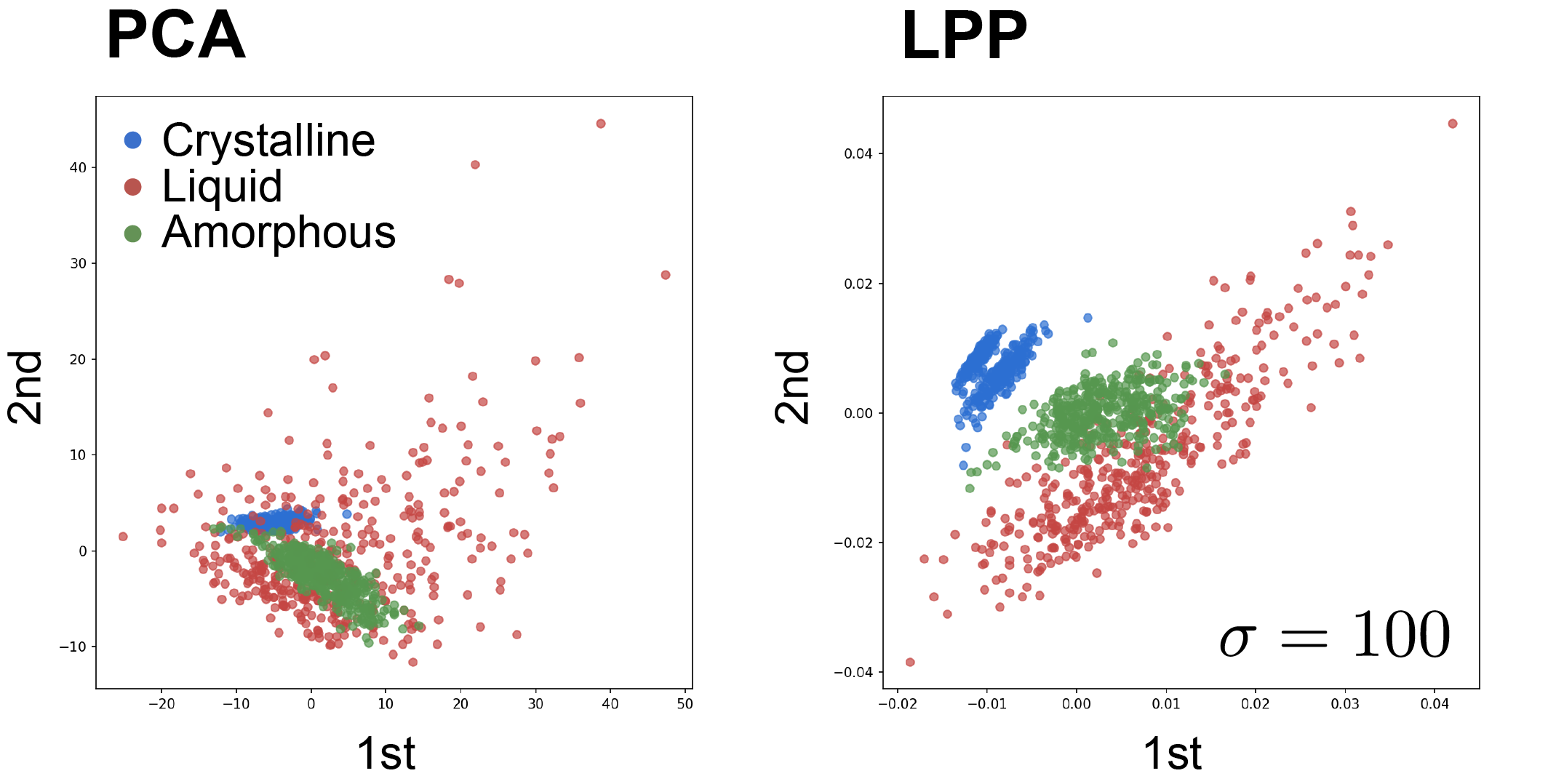}
 \caption{
Distributions after dimension reduction to $d_{\rm r}=2$ by the PCA and LPP methods when $R_{\rm a} = 4 {\rm \AA}$ and $R_{\rm c} = 5 {\rm \AA}$.
Blue, red, and green indicate the crystalline, liquid, and amorphous states, respectively.
The optimal value of $\sigma$ is denoted in the LPP case when the clustering number is fixed at three.
}
\label{pca_lpp}
\end{figure}

To compare the results by other dimensionality reduction methods, where the linear transformation to a low-dimensional space can be obtained,
the distributions by PCA and LPP are shown for the case of $R_{\rm a} = 4 {\rm \AA}$ and $R_{\rm c} = 5 {\rm \AA}$ in Fig.~\ref{pca_lpp}.
The three distributions overlap with each other in both methods.
The PCA method extracts a component (or dimension) showing the largest possible variance under the constraint that it is orthogonal to the preceding components. 
In both 1st and 2nd principal components, the minimum and maximum data points originate from the liquid state. 
In the original LAAF descriptor space, the variation of the data points in the liquid state is probably much larger than that in the crystalline or amorphous state. 
Hence, efficient clustering to distinguish the data points of the amorphous or the crystalline states from the liquid states is difficult. 
Thus, PCA does not seem to work with such a data structure (distributions depending on the cutoff values are shown in Fig.~S2).
On the other hand, if the values of $R_{\rm a}$ and $R_{\rm c}$ increase,
LPP can create a 2D space where the three states are completely divided (distributions depending on the cutoff values are shown in Fig.~S3).
We conclude that TS-LPP is an effective dimensionality reduction method for distinguishing atoms by local structures when $R_{\rm a}$ and $R_{\rm c}$ are sufficiently small.

In addition, the results obtained by other conventional dimensionality reduction methods are summarized in Fig.~S4,
and it is found that these methods are not as efficient as TS-LPP.
Recently, several reports have suggested that two-step analysis for dimension reduction is useful, such as PCA-LPP, PCA-t-SNE, and PCA-UMAP~\cite{lin_identification_2015,lu_hierarchical_2018,kobak_art_2019,sakaue_dimensionality_2020}. 
Our results show that there is a useful case even if the same dimensionality reduction method is repeated in the two-step analysis, although the different algorithms are usually combined.
In Sec.~\ref{discussion_TS-LPP}, 
we will discuss why the TS-LPP can create a better 2D space than PCA and LPP.

%%%%
\subsubsection{Performance for out-of-sample points}

\begin{figure}[t]
\centering
\includegraphics[scale=0.8]{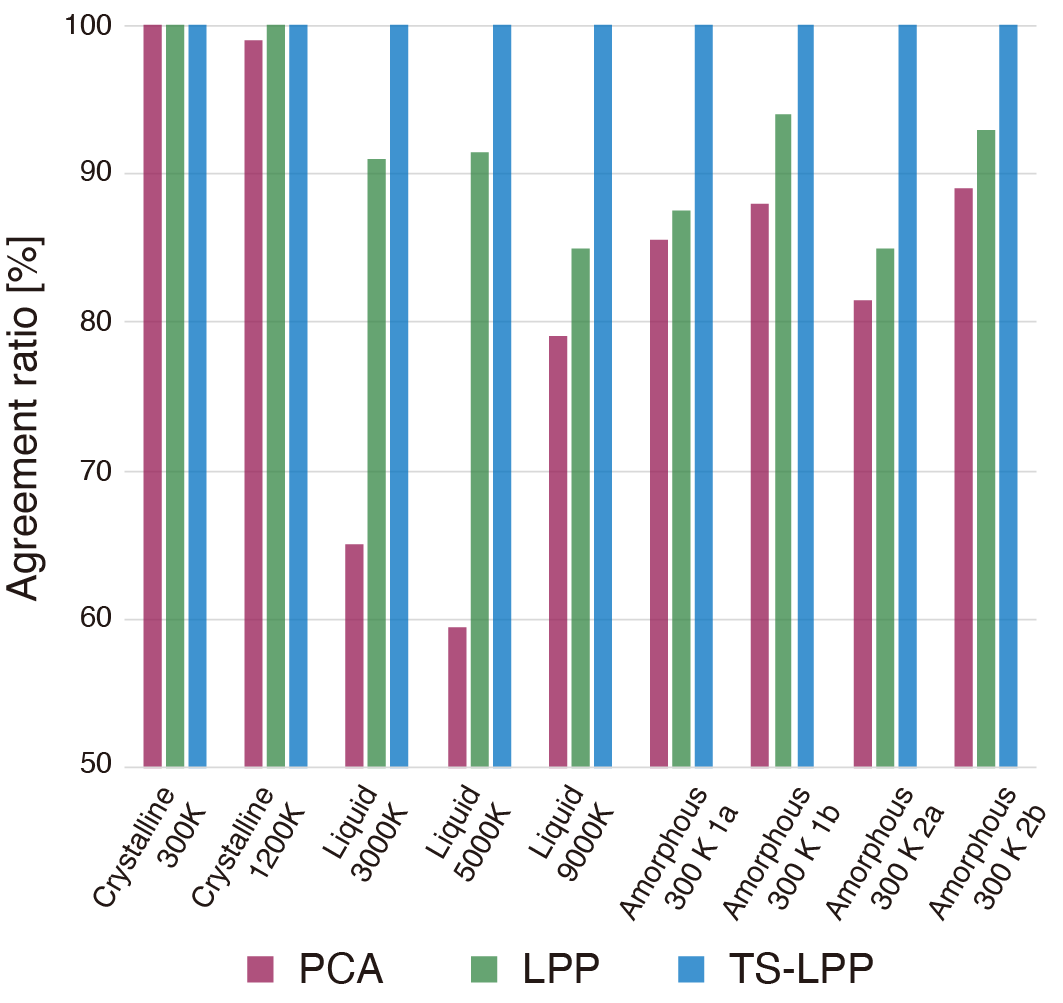}
 \caption{
Ratio of the agreement between the predicted and correct states for new data points by PCA, LPP, and TS-LPP for $R_{\rm a}=4 {\rm \AA}$ and $R_{\rm c}=5 {\rm \AA}$ by the k-nearest neighbor method.
The 2D spaces shown in Figs.~\ref{2d_three_states} (a) and \ref{pca_lpp} are used.
Datapoints of the liquid state at 5000 K and the amorphous states 2a and 2b at 300 K are not included in the training dataset.
}
\label{prediction_single}
\end{figure}

For the proposed method to be capable of various applications, 
it is also important to locate new data points (out-of-sample points), which are not included in the dataset used to create the low-dimensional space. 
Note that some dimensionality reduction methods, such as neural networks and t-SNE, cannot be applied to a new dataset using a linear mapping function. 
For this purpose, we prepare a new dataset from nine FPMD simulations and plot the new data points in the low-dimensional spaces shown in Figs.~\ref{2d_three_states} (a) and \ref{pca_lpp} for $d_{\rm r}=2$, using the linear mapping function obtained by the PCA, LPP, and TS-LPP methods when $R_{\rm a} = 4 {\rm \AA}$ and $R_{\rm c} = 5 {\rm \AA}$. 
The new dataset contains 200 data points for each MD trajectory. 
To evaluate the relationship between the new data and its correct state, we assign each new data point into the crystalline, liquid, or amorphous state, by the k-nearest neighbor method with $k=10$ using the datasets in the low-dimensional spaces.
Then, for each FPMD simulation, we count the number of test data assigned to the correct state in the low-dimensional space. 
Figure~\ref{prediction_single} shows the ratio of the agreement in the case of $d_{\rm r}=2$. 
The classification using PCA does not agree well with the correct state, especially for the liquid states. 
The results of LPP are much better than those of PCA for liquid states, but the results for the amorphous states by these two methods are different, and they are both not reliable. 
In contrast, the TS-LPP results are perfect for all cases, both for the data from the MD trajectories used for creating the low-dimensional space and for those from other trajectories (i.e., liquid at 5000 K, amorphous 2a and 2b at 300 K). 
This result suggests that the reduced dimensions prepared by TS-LPP are transferable and useful, even for the analysis of new data.

%%%%
\subsubsection{\label{discussion_TS-LPP} Advantage of TS-LPP method}

\begin{figure}[t]
\centering
\includegraphics[scale=0.4]{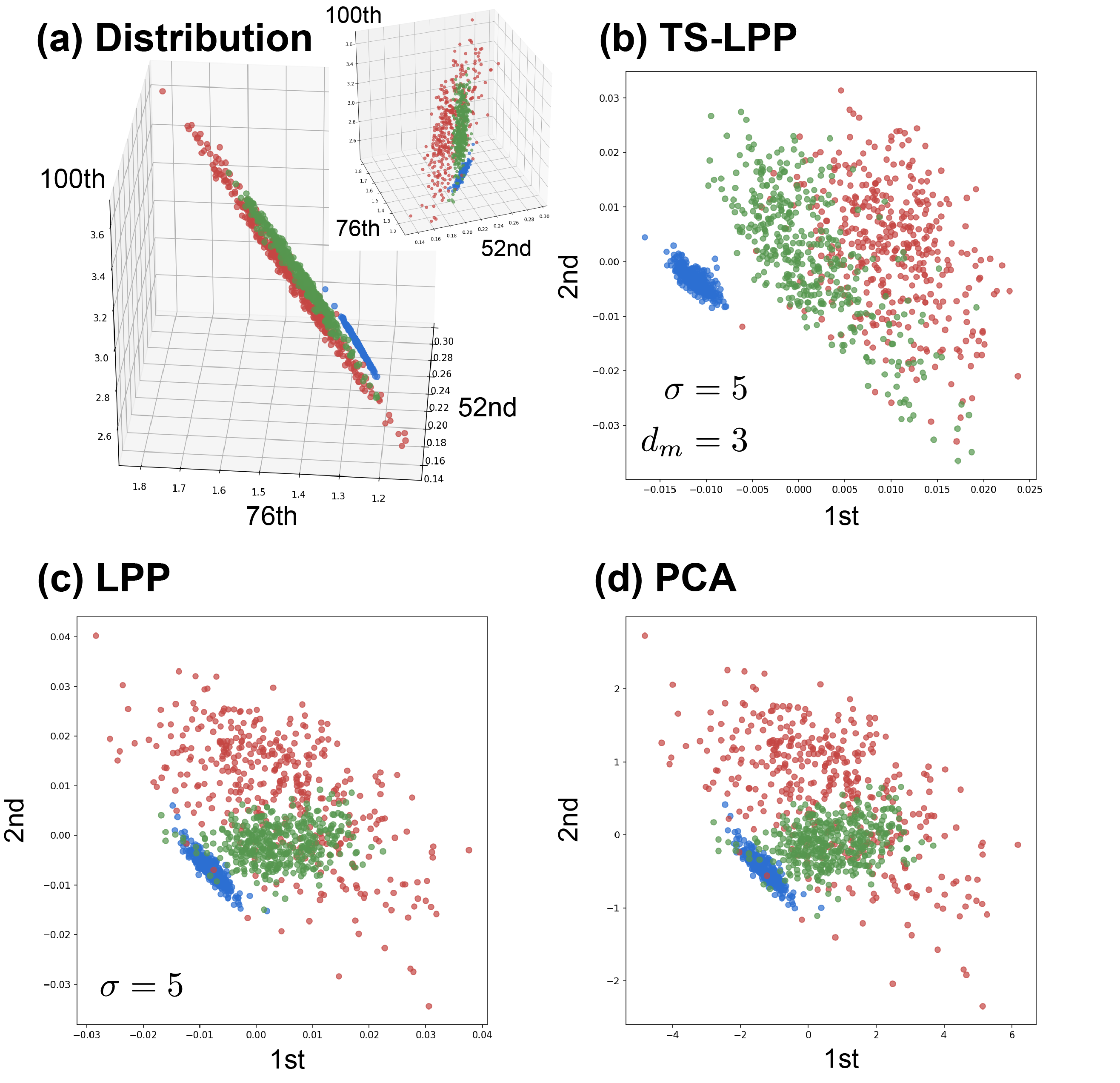}
\caption{
(a) Distributions of the selected three features of 52nd, 76th, and 100th LAAF descriptors when $R_{\rm a} = 4 {\rm \AA}$ and $R_{\rm c} = 5 {\rm \AA}$.
Blue, red, and green indicate the crystalline, liquid, and amorphous states, respectively.
The data points of each state are distributed on a plane.
The orientations of the normal vectors of the planes look similar and the difference of the positions of the planes are small.
Distributions in 2D space by (b) TS-LPP, (c) LPP, and (d) PCA are plotted.
The optimal values of $\sigma$ and $d_{\rm m}$ are denoted when the clustering number is fixed at three.
For this dataset, TS-LPP is better for distinguishing crystalline states from liquid and amorphous states.
}
\label{2D_discuss_real}
\end{figure}

In this section, we discuss why TS-LPP can find a better low-dimensional space for our structural analysis than PCA and LPP.
To visualize the transformation of data distribution performed by dimensionality reduction methods, we perform a 3D to 2D dimensionality reduction.
By a preliminary experiment, we identify and select three dominant descriptors from the 100-dimensional LAAF descriptor that provide a 3D space well-representing the characteristic of the original 100-dimensional space.
A 3D plot for the three descriptors is shown in Fig.~\ref{2D_discuss_real} (a).
For each state (crystalline, liquid, and amorphous), the data points are distributed on a plane.
The orientations of the planes look similar and the difference of the positions of the planes is subtle.
For this 3D dataset, PCA, LPP, and TS-LPP are performed to obtain the 2D spaces, and the results are shown in Figs.~\ref{2D_discuss_real} (b)--(d).
Only TS-LPP can separate the crystalline state from the liquid and amorphous states.

As we have seen in the analyses of synthetic data shown in Sec.~\ref{sec:synthetic_TS-LPP}, 
TS-LPP performs well when the data structure consists of multiple widely distributed data in a low-dimensional space. 
The data given by the present MD simulations has similar data structure (Fig.~\ref{2D_discuss_real}(a)). 
Such data structure will appear, for example, when each feature in the high dimension is not independent but is correlated with each other. 
That is, there is a lower-dimensional space that represents the correlations in the features,
and the data points will be widely distributed within this space. 
In this case, the differences between the classes appear as differences in this lower-dimensional space, 
and the data structure will be a pile of widely distributed data. 
In fact, since the LAAF descriptors used in this study are highly related to each other, 
TS-LPP would be effective for our target data sets. 
Because the descriptors used in materials science and condensed matter physics are often related to each other, 
TS-LPP is an effective method for dimensionality reduction in these fields.

%%%%%%%%%
\subsection{\label{sec:melt-quench} Structural analysis of melt-quench process in Si system}

We analyze the trajectories of the melt-quench process in the Si system are analyzed.
From Sec.~\ref{sec:Si_single}, 
we have two trajectories, that is, melt-quench processes 1 and 2, by FPMD simulations for melt-quench processes (see Fig.~\ref{process_amo}),
and we show the structural analysis results using these trajectories.

\subsubsection{Classification of atoms in melt-quench process}

We apply the method, TS-LPP with the LAAF descriptor, to the analysis of the FPMD simulation of the melt-quench process for amorphous formation. 
Here, we classify all atoms at all snapshots in melt-quench process 1 (see Fig.~\ref{process_amo}) by using the k-nearest neighbor method with $k=10$ in the 2D space obtained by TS-LPP for $R_{\rm a}=4 {\rm \AA}$ and $R_{\rm c}=5 {\rm \AA}$, as shown in Fig.~\ref{2d_three_states} (a).

\begin{figure*}[t]
\centering
\includegraphics[scale=0.35]{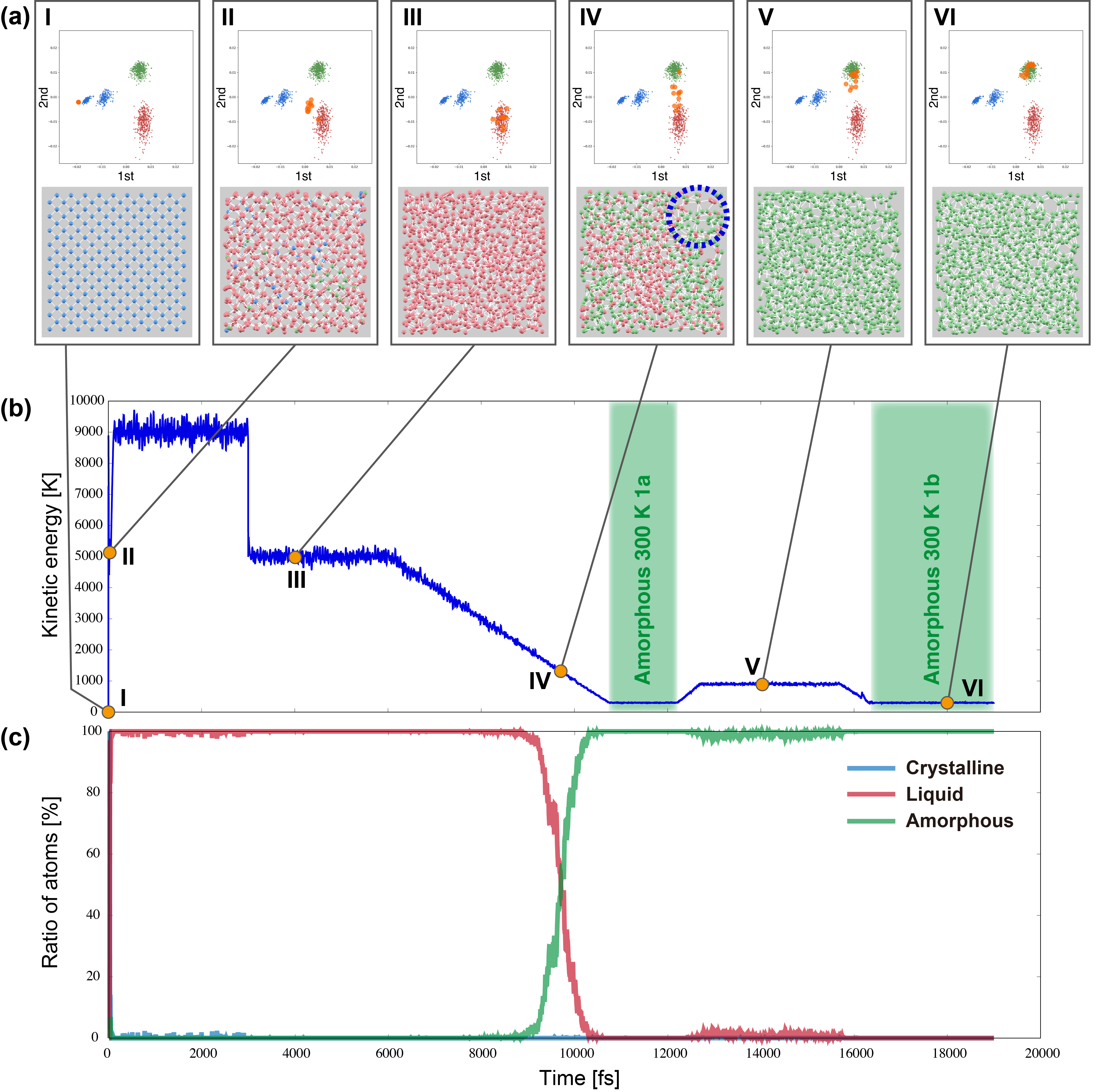}
 \caption{
(a) Data points for the 10 randomly selected atoms in the 2D space by TS-LPP with $R_{\rm a}=4 {\rm \AA}$ and $R_{\rm c}=5 {\rm \AA}$, which are colored in orange (upper figures), and characteristic snapshots (lower figures) in the melt-quench process. 
In the upper figures, the blue, red, and green points are regarded as crystalline, liquid, and amorphous states in the target dataset, respectively. 
In the lower figures, to classify each atom, the k-nearest neighbor method in the 2D space by TS-LPP is used. 
(b) Kinetic energy corresponding to the temperature of the system as a function of time. 
(c) Percentage of atoms belonging to each state.
}
\label{snapshot_class}
\end{figure*}

Figure~\ref{snapshot_class} (a) summarizes the characteristic snapshot structures during the melt-quench process (lower figure) and the data points colored in orange for 10 randomly selected atoms from each snapshot in the 2D space by TS-LPP, together with the points of the dataset used to create low-dimensional space (upper figure). 
The blue, red, and green points denote the crystalline, liquid, and amorphous states, respectively. 
Figure~\ref{snapshot_class} (b) and (c) show the time profile of the temperature and the percentage of atoms belonging to each state, which is predicted by the k-nearest neighbor method, respectively. 
Supplementary Movie 1 shows the time dependence of these results for the entire melt-quench process.

During the melt-quench process, the target temperature is sometimes set as constant for a few picoseconds. 
The system shows an almost equilibrated state during this period, for example, cases III and VI in Fig.~\ref{snapshot_class}.  
In such cases, almost all atoms are classified as the corresponding state, indicating that the data points of all atoms are located near or in the region of the corresponding cluster in the 2D space obtained by TS-LPP. 
In contrast, when the system is cooled or heated, the classified states are mixed. 
For instance, when the temperature rapidly increases from the crystalline state (initial state, case I in Fig.~\ref{snapshot_class}), the three states are mixed (see case II in Fig.~\ref{snapshot_class}). 
In the snapshot of case IV, where the percentages of atoms classified as the liquid state and amorphous state are approximately the same (i.e., around 1280 K), atoms classified as the amorphous state are centered at the upper right part (dotted circle). 
The vacancy seems to act as a nucleus for the phase transition between the liquid and amorphous states. 

More interestingly, in the cooling process from 5000 K to 300 K, the selected 10 atoms are observed to move along the line (mainly along the 2nd axis, i.e., vertical axis) from the liquid to amorphous regions. 
This suggests that the line connecting the two states can be considered as a reaction coordinate for the ``local'' liquid-amorphous transition. 
In case V at 900 K, most atoms are classified as being in the amorphous state, but some atoms appear in the liquid state. 
Such detailed information can be found in the process of finding a more stable amorphous state (i.e., amorphous state 1b at 300 K) during the annealing process. 
These results demonstrate that by using our unsupervised method, we can visually understand the FPMD simulation results in detail.
Note that such analysis would not be possible with PCA,
because new data is often misclassified as shown in Fig.~\ref{prediction_single}.
For melt-quench process 2, the structural analysis of amorphous formation is also performed using the proposed method. 
The results are summarized in the Supplementary Movie 2. 

\subsubsection{Detailed analysis of different amorphous structures}

So far, we have investigated the clustering of the data points from three clearly different groups,
crystalline, liquid, and amorphous states.
Here, we focus on the amorphous phase and perform the detailed analysis
of the local atomic structures in various amorphous states.
For this purpose, we analyze the five amorphous states:
1a, 1b, 2a, and 2b, which are made from the melt-quench process shown in Fig.~\ref{process_amo},
and the one made from the continuous random network model (CRN),
which is an idealized model for perfectly coordinated amorphous semiconductors~\cite{zachariasen_atomic_1932,wooten_computer_1985}.
The quality of the amorphous structure generated by melt-quench methods strongly depends on process, especially on the quenching speed~\cite{Deringer-2018}.
It is found that some detailed feature of the experimental structure factor cannot be reproduced by the amorphous structures made by the present melt-quench processes.
On the other hand, the fifth structure is generated from a snapshot of the NPT simulation starting from the high-quality CRN structure previously reported, which agrees well with experimental results~\cite{Barkema-2000}.

\begin{figure}[t]
\centering
\includegraphics[scale=0.4]{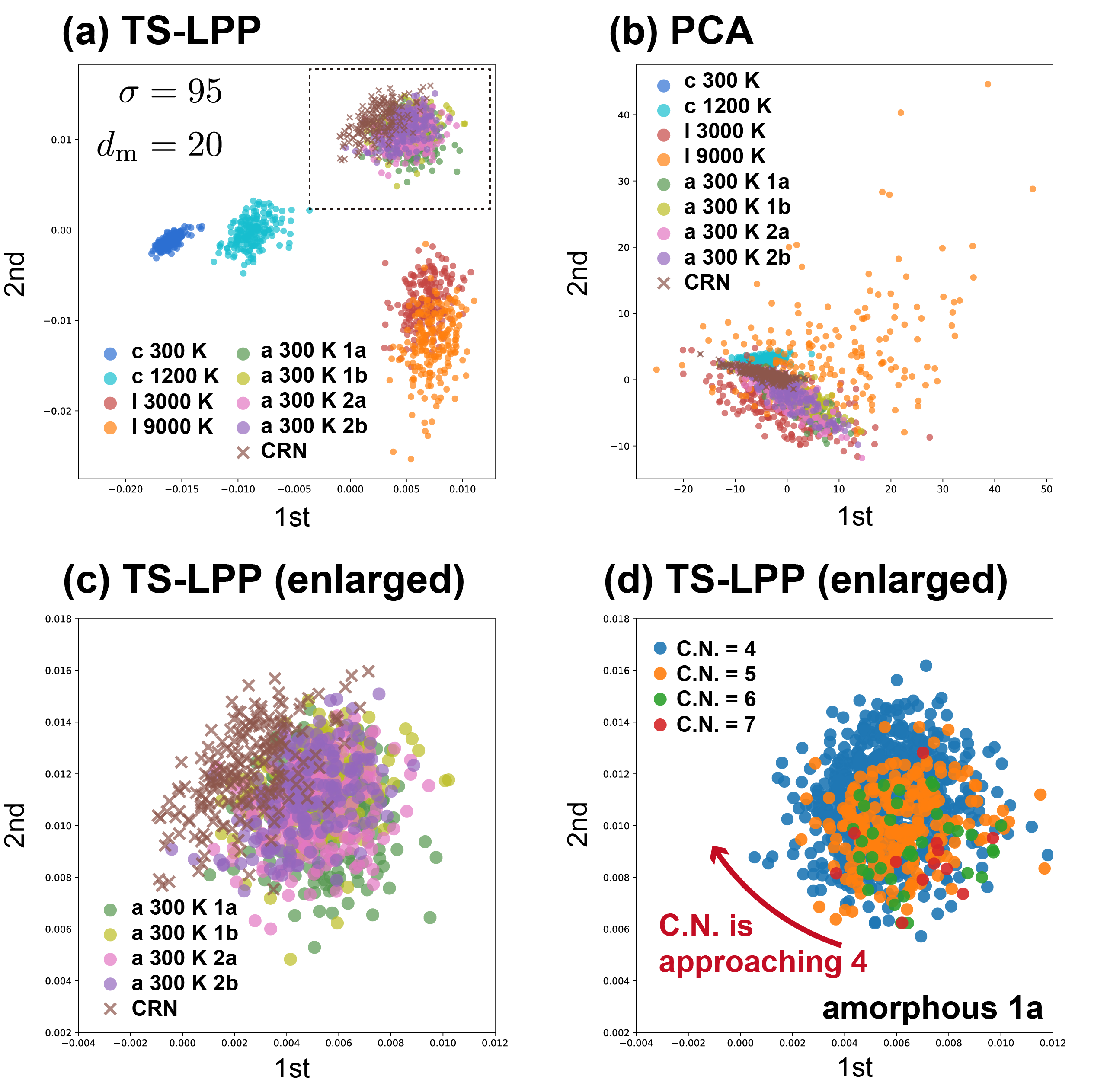}
 \caption{
Distributions of datapoints for amorphous states 2a and 2b at 300 K, and the CRN results mapped to the 2D space constructed by (a) TS-LPP method and (b) PCA, when $R_{\rm a}=4 {\rm \AA}$ and $R_{\rm c}=5 {\rm \AA}$.
(c) Enlarged view of the amorphous states of TS-LPP.
The region is denoted by the dotted line in (a).
(d) Distributions of datapoints for 1000 atoms in amorphous 1a depending on the coordination number (C.N.).
The region is denoted by the dotted line in (a).
}
\label{amorphous_analysis}
\end{figure}

The data points sampled from these five amorphous states are mapped to
the low-dimensional spaces obtained by the TS-LPP method (Fig.~\ref{2d_three_states} (a)) and
the PCA method (Fig.~\ref{pca_lpp}) when $R_{\rm a}=4 {\rm \AA}$ and $R_{\rm c}=5 {\rm \AA}$.
The distributions of 200 data points for each amorphous state are shown in
Figs.~\ref{amorphous_analysis}(a) and (b), and an enlarged view of TS-LPP results is shown in Fig.~\ref{amorphous_analysis}(c).
Note that, different from Figs.~\ref{2d_three_states} and \ref{pca_lpp}, the same phase with different temperatures are distinguished by different colors.
In the low dimensional space created by PCA (Fig.~\ref{amorphous_analysis}(b)), the distribution of the data points sampled from the crystalline, liquid and the five amorphous states are all overlapped.
Although the distribution of the CRN model is localized in the space by PCA,
it is difficult to extract its unique structural properties from those of other states.

On the other hand, in the space created by TS-LPP, the distribution of the three phases are well separated even with the newly added amorphous data points.
From Fig.~\ref{amorphous_analysis}(a), it can be seen that the new data points sampled from the amorphous 2a and 2b states are distributed close to the results of 1a and 1b in both cases, which is consistent with the prediction of states using the k-nearest neighbor method (see Fig.~\ref{prediction_single}).
The data points from the CRN model are also distributed close to the region of 1a and 1b, 
but it is found that the center of its distribution is slightly shifted to upper left from those of 1a, 1b, 2a, and 2b.
Furthermore, as can be seen in Fig.~\ref{amorphous_analysis}(c), the distribution of 1b is closer to the region of CRN model than 1a. 
This tendency is observed also in the data points from the melt-quench process 2; the distribution of 2b is closer to the one for the CRN model than 2a.
It supports that the annealing process to 900K (see Fig.~\ref{process_amo}) is effective for the modeling of the amorphous structure.
Furthermore, we also analyze the coordination number of the data points of amorphous state 1a, whose distribution is most deviated from that of CRN.
The mapping of all 1000 atoms in one snapshot of amorphous state 1a is shown in Fig.~\ref{amorphous_analysis}(d), 
where the color differences represent the coordination number of each atom.
Note that all of the atoms are four coordinated in the CRN model.
While most of the atoms in the 1a state are also four coordinated, there are some atoms whose coordination number is larger than 4.
It is found that such atoms tend to distribute in the lower right region, deviated from the center of the distribution of the CRN model.
Considering that the number of atoms in this region decrease by the annealing process, one of the effects by annealing process is to reduce the highly coordinated atoms.
However, it should be noted that there are also many atoms whose coordination number is 4 in the same lower right region.
Thus, the low-dimensional space constructed by the TS-LPP method is not simply related to the coordination number and we expect the present analysis would be useful to extract more detailed information in the structural change during the annealing process.
Such attempts are now under way, and further reports will be presented in the future.

\subsubsection{Analysis of the process from liquid to amorphous formation}

\begin{figure}[t]
\centering
\includegraphics[scale=0.4]{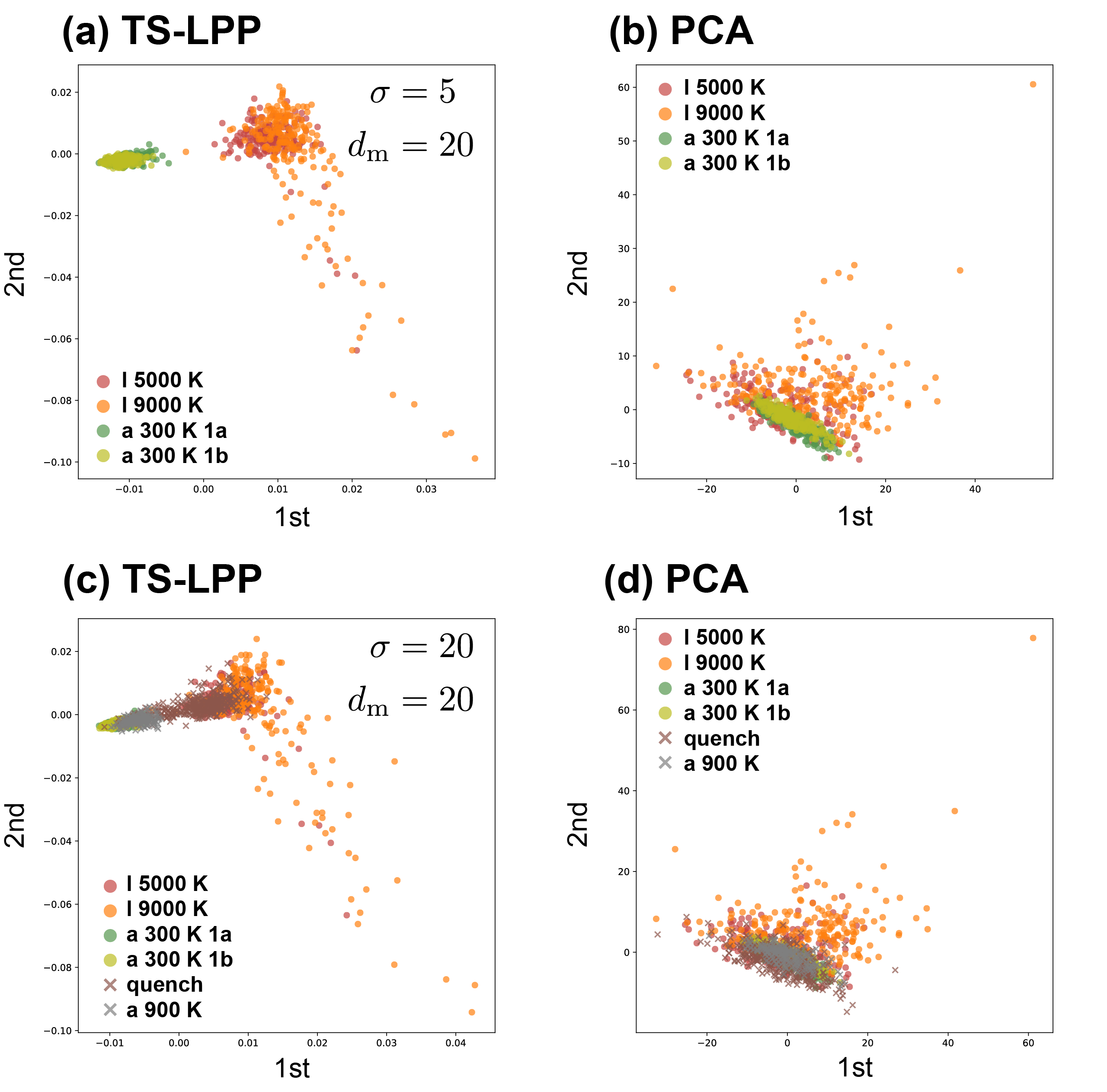}
 \caption{
Distributions of datapoints for  liquid states at 5000 K and 9000 K, amorphous states 1a and 1b at 300 K in the low-dimensional space constructed by (a) TS-LPP and (b) PCA, when $R_{\rm a}=4 {\rm \AA}$ and $R_{\rm c}=5 {\rm \AA}$.
Distributions of datapoints for quench states and amorphous states at 900 K in addition to the cases in (a) and (b) by (c) TS-LPP and (d) PCA.
}
\label{melt_quench_analysis}
\end{figure}

So far, we have searched for low-dimensional spaces where crystalline, liquid, and amorphous states are well distinguished. 
Here, we address the structural analysis when there are only liquid and amorphous states. 
Figures~\ref{melt_quench_analysis} (a) and (b) show the 2D space obtained by TS-LPP method and PCA, respectively, 
when liquid states at 5000 K and 9000 K and amorphous states 1a and 1b at 300 K are adopted.
In TS-LPP, it can be seen that the liquid and amorphous states can be separated on the 1st axis. 
Furthermore, the amorphous state 1a at 300 K is closer to the liquid states than to 1b. 
Although the number of clusters for determining hyperparameters is set to three, it is confirmed that even if the cluster number is increased, the obtained results are almost the same. 
In contrast, when using PCA, all the results overlap, 
and it is difficult to extract the relations of atoms with each other.

Next, we generate a dataset from the entire melt-quench process and consider the low-dimensional space created by the entire dataset of the melt-quench process (see Fig.~\ref{process_amo}). 
The dataset includes 200 points each from liquid states at 5000 K, 9000 K, amorphous states 1a and 1b at 300 K, and amorphous state at 900 K. 
In addition, 400 points during the quench process from liquid state at 5000 K to amorphous state 1a at 300 K are also included. 
The results of the low-dimensional space obtained by the TS-LPP method and PCA are shown in Figs.~\ref{melt_quench_analysis} (c) and (d). 
In this way, by the TS-LPP method, 
the liquid and amorphous states are separated even when the quench state is added, 
and the quench state is distributed to connect the liquid and amorphous states. 
Thus, we can consider that the 1st axis is apparently related to the temperature. 
In contrast, in the PCA results, all the data are overlapped. 
Therefore, it is shown that by using TS-LPP, 
we can find an appropriate low-dimensional space even if the entire melt-quench process is included in the dataset.

%%%%%%%%%
\subsection{\label{sec:SiGe_binary} Structural analysis of crystalline, liquid, and amorphous states in Si-Ge system}

To confirm that our method is widely usable,
we perform structural analysis for a silicon-germanium binary system in the crystalline, liquid, and amorphous phases, which are prepared by FPMD simulations using CONQUEST. 
To prepare crystalline and liquid phases, a system containing 500 silicon and 500 germanium atoms at a constant temperature of 300 K for crystalline and 5000 K for liquid and constant volume is simulated. 
The alloy where silicon and germanium atoms are arranged alternately in the lattice structure is considered as a crystalline state. 
On the other hand, the amorphous structures are simulated by a system with 550 silicon and 450 germanium atoms at 300 K prepared by the melt-quench process 2, as shown in Fig.~\ref{process_amo}.

\begin{figure}[t]
\centering
\includegraphics[scale=0.4]{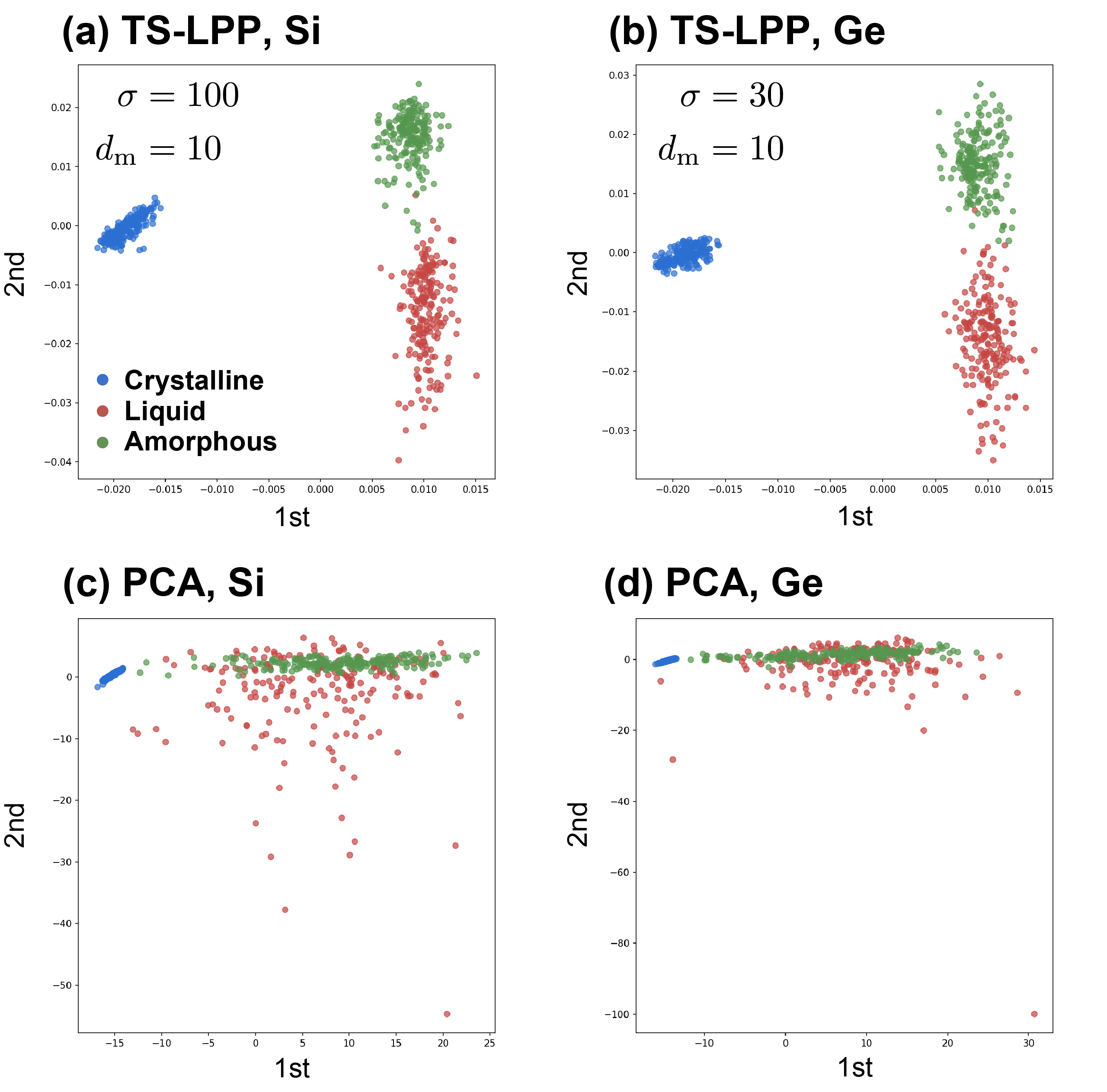}
 \caption{
Distributions after dimension reduction to $d_{\rm r} =2$ by the TS-LPP method for (a) silicon and (b) germanium. 
Distributions obtained by PCA for (c) silicon and (d) germanium.
LAAF descriptors for binary systems with $R_{\rm a}=4 {\rm \AA}$ and $R_{\rm c}=5 {\rm \AA}$ are used.
}
\label{binary_analysis}
\end{figure}

Figure~\ref{binary_analysis} shows the distributions of datapoints for silicon and germanium atoms in the 2D space ($d_{\rm r}=2$) by TS-LPP and PCA when $R_{\rm a}=4 {\rm \AA}$ and $R_{\rm c}=5 {\rm \AA}$.
For TS-LPP, the number of clusters for determining hyperparameters is set to three, but the cluster number is not an important factor. 
In the case of TS-LPP, the three states are well separated in both silicon and germanium.
The 1st axis divides the crystalline states from the liquid and amorphous states, while the liquid and amorphous states are separated by the second axis.
This is the same result obtained in silicon single-component system (see Fig.~\ref{2d_three_states}).
On the other hand, in the case of PCA, the liquid and amorphous states completely overlap, and the crystalline states are located near other states.
Thus, we conclude that even if a binary system is considered in a straightforward manner, the TS-LPP method can find a well-defined low-dimensional space, where the distributions of datapoints for each atom or group of atoms can be properly captured.
This result indicates that the method has the potential to be applied to more complex systems.

%%%%%%%%%
\subsection{\label{sec:Cu_single} Structure analysis for Cu system}

So far, only covalent bond systems are considered.
We next analyze the structures of metallic systems, fcc, hcp and liquid states of the single component copper system. 
Note that fcc and hcp are both close-packed structures and it is difficult to distinguish them from the present LAAF descriptor, 
which includes only the radial term without the angle information of three atoms.
We investigate the ability of TS-LPP method for this analysis.
To generate the local structures of the copper system, 
classical MD simulations with the EMT potential are performed using the Atomic
Simulation Environment (ASE) package~\cite{larsen_atomic_2017}. 
The number of atoms is 500. 
MD calculations are performed at a constant temperature of 300 K (fcc and hcp crystalline phases) and 5000 K (liquid) and constant volume.
Data is generated by extracting 200 points from each state, 
and structural analysis is performed using 600 data points in total.

\begin{figure}[t]
\centering
\includegraphics[scale=0.4]{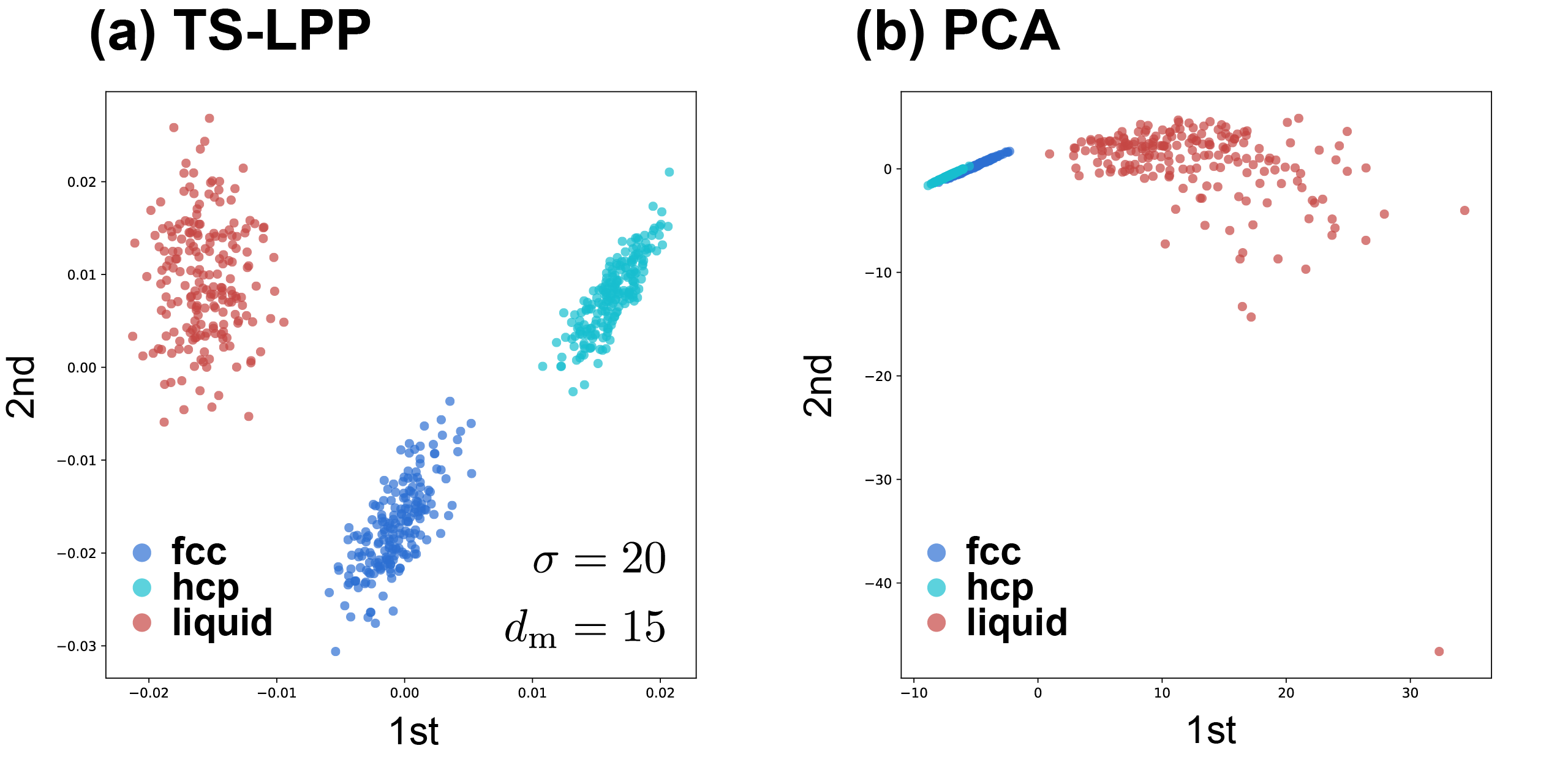}
 \caption{
Distributions in 2D space obtained by (a) TS-LPP and (b) PCA for $R_{\rm a}=3 {\rm \AA}$ and $R_{\rm c}=3 {\rm \AA}$ for Cu system.
The fcc and hcp structures are prepared as a crystalline states. 
}
\label{copper_analysis}
\end{figure}

The obtained 2D spaces for $R_{\rm a}=3 {\rm \AA}$ and $R_{\rm c}=3 {\rm \AA}$ are shown in Figs.~\ref{copper_analysis}(a) and (b) when TS-LPP and PCA are used. 
For TS-LPP, the number of clusters for determining hyperparameters is set to three, although we confirm that the distributions are almost the same even when the cluster number is changed. 
In both TS-LPP and PCA, it can be seen that the crystalline 300 K and liquid 5000 K states are separated. 
The fcc and hcp data can be distinguished by TS-LPP,
while these data are overlapped in PCA result.

%%%%%%%%%%%%%%%%%%%%%%%
%%%%%%%%%%%%%%%%%%%%%%%
\section{\label{sec:summary}Discussion and summary}
%%%%%%%%%%%%%%%%%%%%%%%

In summary, we have proposed a novel structural analysis method based on an unsupervised ML technique and applied it to structural analysis problems.
Our method can find a low-dimensional space where the distributions of datapoints for each atom or group of atoms can be properly captured using the proposed LAAF descriptor showing the local structure without labels. 
The key to our method is to perform dimensionality reduction from a high-dimensional descriptor space to a lower dimension using TS-LPP, where conventional LPP is performed twice. 
Our method is much more reliable than PCA or other conventional dimensionality reduction methods. 
Furthermore, our method is a parameter-free unsupervised technique, since the hyperparameters in the method can be determined automatically. 
Our implementation of the TS-LPP method is available on GitHub~\cite{noauthor_httpsgithubcomrtmrts-lpp_2020}. 
We have used the method to analyze the FPMD simulations of a silicon single-component system,
silicon and germanium binary system, and classical MD simulations for copper single-component system.
Through these applications, we have demonstrated that the method is effective for understanding the phenomena in detail and for elucidating the local phenomena during the simulations.

We believe that the new method is useful in many kinds of applications. 
As the method is unsupervised, it can be used to analyze the structures of unknown phases or materials. 
In addition, we can control the resolution of the structural analysis depending on the degree of dimension. 
Here, we discuss two possible applications. 
The first application is the development of highly accurate and transferable ML force fields. 
In many cases, the problem of the ML forces is the transferability~\cite{kuritz_size_2018,tamura_machine_2019}. 
If we can construct the force fields from multiple ML forces based on different sets of training data, depending on the local environment, the accuracy of ML forces can be improved easily. 
The classification method proposed in this study should be effective for this purpose as well as the Gaussian mixture model~\cite{pham_novel_2016}. 
For example, in many systems that include irregular regions, such as surface or interface structures, there are structural variations, and it is difficult to determine how the atoms in different regions should be classified. 
Even in such cases, since our method is unsupervised, we can simply collect the data from a short FPMD simulation and classify the atoms into several groups depending on the data structure. 
If we construct the ML forces based on this classification, the accuracy of the ML forces should be high. 
The proposed method can also detect a new structures during a long MD simulation using the constructed ML forces. 
Therefore, on-the-fly development of ML forces of large, irregular, and complex systems should be possible with a large-scale DFT code, such as the CONQUEST code that we used in this work.

The second application is to find a useful coordinate to characterize the local phenomena in complex dynamical processes, chemical reactions, or phase transformations. 
As we have pointed out in the analysis of the melt-quench process, the component in the low-dimensional space can be considered as a good reaction coordinate to express the ``local'' transformation from a liquid to an amorphous state. 
Note that although the present descriptor is ``local'' and specific to each atom, the descriptor also includes the information of its neighbors. 
With the LAAF descriptor, we can follow the phenomenon of a group of atoms. 
If TS-LPP with the LAAF descriptor can find useful coordinates for many kinds of phenomena, we expect that it can be helpful in calculating the energy barrier for the local reactions or for the efficient structural search of new phases. 
We intend to explore these two applications in future work.

\begin{acknowledgments}
We thank Ayako Nakata and Masato Shimono for the useful discussions. 
J.L. and T.M. are grateful to Zamaan Raza for his help on the MD simulations of amorphous Si using CONQUEST, and Normand Mousseau for providing the CRN amorphous Si models.
This article is based on the results obtained from a project subsidized by the ``Materials Research by Information Integration'' Initiative (MI2I) project, Core Research for Evolutional Science and Technology (CREST) (Grant No. JPMJCR17J2) from the Japan Science and Technology Agency (JST), the New Energy and Industrial Technology Development Organization of Japan (NEDO) Grant (JPNP16010, AJD30064), and JSPS Grant-in-Aid for Scientific Research (Grant Numbers 18H01143, 18H03250, 19K20280, and 21H01008). 
This work was partly supported by the World Premier International Research Centre Initiative (WPI Initiative) on Materials Nanoarchitectonics (MANA) and the Multidisciplinary Cooperative Research Program in CCS, University of Tsukuba. 
The computations in the present work were performed on the Numerical Materials Simulator at NIMS and the supercomputer at the Supercomputer Center, Institute for Crystalline State Physics, The University of Tokyo.
\end{acknowledgments}

\appendix

\section{Low-rank approximation for LPP}

For our target data,
the input data matrix $X$ has a number of small nonzero singular values.
Thus, to improve the numerical stability for calculations in LPP, 
we use the low-rank approximation of $X$ using the singular value decomposition $U \Sigma V^\top = X$. 
Here, $U$ is $N \times M$, the columns of which are the left singular vectors of $X$. 
$\Sigma=\text{diag}(s_1,s_2,... ,s_M)$ is an $M \times M$ diagonal matrix with the descending ordered singular values $\{s_i \}_{i=1,...,M}$ of $X$. 
$V$ is $N \times M$ matrix, the columns of which are the right singular vectors of $X$. 
By introducing a parameter $0 < \delta <1 $, the numerical rank of $\rho$ is determined by: 
\begin{eqnarray}
\frac{s_{\rho+1}}{s_1} < \delta < \frac{s_\rho}{s_1}.
\end{eqnarray}
Let $U_0$ and $V_0$ be matrices with the $\rho$ leftmost columns of $U$ and $V$, respectively, and $\Sigma_0$ be the leading principal submatrix of order $\rho$ of $\Sigma$. 
Using these matrices, instead of solving Eq.~(\ref{eq:LPP}), we solve the generalized eigenvalue problem as: 
\begin{eqnarray}
U_0^\top L U_0 \mathbf{z} = \lambda U_0^\top D U_0 \mathbf{z},
\end{eqnarray}
where $\{\lambda_i, \mathbf{z}_i \}$ is the $i$th eigenvalue and eigenvector. 
The alternative mapping matrix $Y=(\mathbf{y}_1,\mathbf{y}_2,... ,\mathbf{y}_{d_{\rm r}})$ is obtained by
\begin{eqnarray}
\mathbf{y}_i = V_0 \Sigma_0^{-1} \mathbf{z}_i,  \ \ (i= 1,..., d_{\rm r}).
\end{eqnarray}

%%%%%%%%%%%%%%%%%%%%%%%%%%%%%%%%
%References
%%%%%%%%%%%%%%%%%%%%%%%%%%%%%%%%

%\bibliography{TSLPP}


\end{document}